\documentclass[a4paper,11pt]{article}
\pdfoutput=1 

\usepackage{jheppub} 

\usepackage[T1]{fontenc} 

\usepackage{makecell}
\usepackage{graphicx, caption, subcaption}

\title{\boldmath Michel parameters in the presence of massive Dirac and Majorana neutrinos}

\author[]{Juan Manuel Márquez,}
\author[]{Gabriel López Castro}
\author[]{and Pablo Roig}

\affiliation[]{Departamento de Física, Centro de Investigación y de Estudios Avanzados del Instituto Politécnico Nacional\\Apartado Postal 14-740, 07000 Ciudad de México, México}

\emailAdd{jmarquez@fis.cinvestav.mx}
\emailAdd{glopez@fis.cinvestav.mx}
\emailAdd{proig@fis.cinvestav.mx}

\abstract{We analyse the effects of Dirac and Majorana neutrinos on leptonic decays of muon and tau leptons using the most general four-lepton effective interaction Hamiltonian of dimension six. We calculate the specific energy and angular distribution of the final charged lepton, including the polarizations of the initial and final charged leptons. We discuss the new generalized Michel parameters and focus on the effects of the heavy neutrino masses that would lead to sizable contributions on scenarios where the new sterile neutrinos have non-negligible mixing. Specifically, the most promising scenario is found for the case of $\tau$ decay with one heavy final-state neutrino with a mass around $10^2 - 10^3$ MeV, where the linear term suppression could be of order $10^{-4}$, low enough to be measured in current and forthcoming experiments.
}

\begin{document} 
\maketitle
\flushbottom

\section{Introduction}
\label{sec:intro}
The Standard Model of Elementary Particles (SM) is the most successful theory  describing nature. It has been tested so many times, with ever-increasing precision, and remains consistent with the experimental data \cite{1}. Nevertheless, the SM does not explain or include  everything, such as gravity interaction,  neutrino masses, dark matter, baryon asymmetry, etc. These are some physical phenomena that motivate the search for new physics (NP) that is beyond the SM \cite{2}. \\
Nowadays, the Dirac or Majorana nature of neutrinos as well as the mechanism from which they acquire mass are one of the unsolved puzzles that lie beyond the SM. There are many minimal extensions of it in order to account for nonzero masses and mixings for the active neutrinos; adding new gauge singlet fields, such as right-handed neutrinos. Some of them are the well-known $\nu$SM \cite{3,4}  and the seesaw-mechanisms \cite{5,6,7,8,9}. \\ 
Also, the Dirac or Majorana nature of neutrinos is crucial for understanding the origin of their masses and some physical processes. If neutrinos were Majorana particles, then lepton number is not conserved and the neutrinos would be their own antiparticles, i.e, there is no conserved quantum number that allows to distinguish between neutrino or anti-neutrino. This makes it interesting, from the phenomenological point of view,  to test the Dirac or Majorana nature of the neutrinos through lepton number violating processes, specially in the heavy sector, where the Majorana contribution could lead to measurable changes \cite{10}.\\ 
Much theoretical and experimental researches are being carried out to discover how the nature of neutrinos and the specific origin of their masses can affect physical processes, leading to small deviations from the well-known SM results and new decay modes \cite{5,11,12,13,14}.\\
One of the experimental strategies in the search for a fundamental description of nature which goes beyond the SM is to perform high precision measurements where an observed discrepancy with the SM would reveal the signature of NP. For this purpose, muon and tau leptons are specially important, where many precision measurements can be made, see e.g. refs.~\cite{15,16,17}.\\
For example, studies of muon decays both determine the overall strength and establish the Lorentz structure
of weak interactions, as well as setting extraordinary limits on charged lepton
flavor violating processes (LFV) \cite{18}. Measurements of the muon’s anomalous magnetic
moment \cite{Muong-2:2006rrc, Muong-2:2021ojo} offer singular sensitivity to the completeness of the standard model
\cite{Aoyama:2020ynm,1} and the predictions of many speculative theories. In general, due to the amazing high-precision
and extreme-sensitivity  experiments with muons, this lepton is a perfect candidate for precision SM tests.\\
On the other hand, the tau lepton experiments are not as accurate as those made with muons. However, the tau lepton has some unique features that make it as important as the muon in the search for NP. Nowadays, the detailed study of higher-order electroweak corrections and QCD contributions has promoted the physics of the tau lepton to the level of precision tests. The pure leptonic or semileptonic character of tau decays provides a clean laboratory to test the structure of the weak currents and the universality of their couplings to the gauge bosons. Moreover, the tau is the only known lepton massive enough to decay into hadrons; its semileptonic decays are then an ideal tool for studying strong interaction effects in very clean conditions \cite{15}.\\
Finally, since one naively expects the fermions to be sensitive to the possible NP proportionally to their masses, the large tau mass allows one to investigate NP contributions, through a broad range of kinematically-allowed decay modes, complementing the high-precision searches performed in muon decay.\\
In this work we analyse the leptonic decays $\ell^{-}\longrightarrow \ell^{'^{-}}\Bar{\nu}_{\ell^{'}}\nu_\ell$, where the lepton pair ($\ell$,$\ell^{'}$) may be ($\mu$,$e$), ($\tau$,$e$) or ($\tau$,$\mu$), using the most general four-lepton effective interaction Hamiltonian to test the structure of the weak currents and the nature of neutrinos, specially in the framework of low-scale seesaw scenarios, where the new sterile 
neutrinos have non-negligible mixings and some of them require masses low enough to be produced on-shell, where the genuine effects of the heavy masses could be measurable.\\
This article is structured as follows: after a brief review of the well-known Michel distribution (massless case) in section \ref{sec:2}, we present the effective decay rate in section \ref{sec:3} taking into account finite neutrino masses of Dirac (section \ref{subsec:3.1}) and Majorana (section \ref{subsec:3.2}) type, summarizing both results in a single expression (section \ref{subsec:3.3}). 
Then, in section \ref{sec:4} we compare both distributions, discussing their common factors (section \ref{subsec:4.1}) as well as their main differences (section \ref{subsec:4.2}), giving some examples where the specific neutrino nature can be tested. Finally, our conclusions are given in section \ref{sec:5}. Appendices include W-boson propagator corrections to the polarized leptonic decay rate (appendix \ref{appendix:A}) as well as a review of the Fierz transformations (appendix \ref{appendix:B}) relating the most commonly used set of Hamiltonians in these lepton decays.

\section{Lorentz Structure of the Charged Current}
\label{sec:2}
Let us consider the leptonic decays $\ell^{-}\longrightarrow \ell^{'^{-}}\Bar{\nu}_{\ell^{'}}\nu_\ell$ with massless neutrinos, where the lepton pair ($\ell$,$\ell^{'}$) may be ($\mu$,$e$), ($\tau$,$e$) or ($\tau$,$\mu$).\\
The most general, local, derivative-free, lepton-number conserving, four-lepton interaction Hamiltonian, consistent with locality and Lorentz invariance is \cite{15,19,Bouchiat:1957zz}
\begin{equation}
\label{eqn:H}
    \mathcal{H}=4\frac{G_{\ell\ell^{'}}}{\sqrt{2}}\sum_{n,\epsilon,\omega} g_{\epsilon\omega}^n \left[\Bar{\ell}^{'}_\epsilon\Gamma^n(\nu_{\ell^{'}})_\sigma\right]\left[(\Bar{\nu_\ell})_\lambda\Gamma_n \ell_\omega\right].
\end{equation}
The subindices $\epsilon,\omega,\sigma,\lambda$ label the chiralities $(L,R)$ --for left- and right-handed, respectively-- of the corresponding fermions, and $n=S,V,T$ the type of interaction: scalar $(\Gamma^S=I)$, vector $(\Gamma^V=\gamma^\mu)$ and tensor $(\Gamma^T=\sigma^{\mu\nu}/\sqrt{2})$. Since tensor interactions can contribute only for opposite chiralities of the charged leptons, this leads to the existence of 10 complex coupling constants, related to 4 scalar, 4 vector and 2 tensor interactions.\\
Once an unphysical global phase is removed, it leaves 19 real numbers to be extracted from  the experiment. Furthermore the global factor $G_{\ell\ell^{'}}$, which is determined from the total decay rate, leads to the following normalization of the coupling constants \cite{20}
\begin{equation}
\label{eqn:normalization}
\begin{split}
    1&=\frac{1}{4}(|g_{RR}^S|^2+|g_{RL}^S|^2+|g_{LR}^S|^2+|g_{LL}^S|^2)+3(|g_{LR}^T|^2+|g_{RL}^T|^2)\\
    &+(|g_{RR}^V|^2+|g_{RL}^V|^2+|g_{LR}^V|^2+|g_{LL}^V|^2).
    \end{split}
\end{equation}
Thus, $|g_{\epsilon\omega}^S|\leq2$,$|g_{\epsilon\omega}^V|\leq1$ and $|g_{\epsilon\omega}^T|\leq1/\sqrt{3}$. The Standard Model predicts $|g_{LL}^V|=1$ and all others couplings vanishing. In the search for new physics it is important to calculate the final charged-lepton distribution using the Hamiltonian (\ref{eqn:H}).\\
Working with this Hamiltonian, considering massless neutrinos, the differential decay probability to obtain a final charged lepton with (reduced) energy between $x= E_{\ell^{'}}/\omega$ and $x+dx$,
emitted in the direction $\hat{z}$ at an angle between $\theta$ and $\theta+d\theta$ with respect to the initial lepton polarization vector $\mathcal{P}$, and with its spin parallel to the arbitrary direction $\hat{\zeta}$, neglecting radiative corrections, 
is given by \cite{1}
\begin{equation}
\label{eqn:finalpolfl}
    \begin{split}
        \frac{d\Gamma}{dx d\cos{\theta}}&=\frac{m_\ell}{4 \pi^3} \omega^4 G_{\ell\ell^{'}}^2\sqrt{x^2-x_0^2}\\
        &\times\Big(F(x)-\frac{\xi}{3}\mathcal{P}\sqrt{x^2-x_0^2}\cos{\theta}A(x)\Big)\\
        &\times\big[1+\hat{\zeta}\cdot \Vec{\mathcal{P}}_{\ell^{'}}(x,\theta)\big],
    \end{split}
\end{equation}
where $w\equiv(m_\ell^2+m_{\ell^{'}}^2)/2m_\ell$, $x_0\equiv m_{\ell^{'}}/\omega$ and the polarization vector $\Vec{\mathcal{P}}_{\ell^{'}}$ in eq.(\ref{eqn:finalpolfl}) is
\begin{equation}
    \Vec{\mathcal{P}}_{\ell^{'}}=P_{T_1}\ \hat{x}+P_{T_2}\ \hat{y}+P_{L}\ \hat{z}.
\end{equation}
Here $\hat{x},\hat{y}$ and $\hat{z}$ are unit vectors defined as:
\begin{equation}
    \begin{split}
        &\hat{z}\quad \text{is along the $\ell^{'}$momentum $\Vec{p}_{\ell^{'}}$,}\\
        \frac{\hat{z}\times\Vec{\mathcal{P}}_{\ell^{'}}}{|\hat{z}\times\Vec{\mathcal{P}}_{\ell^{'}}|}= &\hat{y}\quad \text{is transverse to $\Vec{p}_{\ell^{'}}$ and perpendicular to the decay plane,}\\
        \hat{y}\times\hat{z}=& \hat{x}\quad \text{is transverse to $\Vec{p}_{\ell^{'}}$ and in the decay plane,}
    \end{split}
\end{equation}
and the components of $\Vec{\mathcal{P}}_{\ell^{'}}$ are, respectively
\begin{equation}
    \begin{split}
        &P_{T_1}=\mathcal{P}\sin{\theta}\cdot F_{T_1}(x)\ /\ \Big\{F(x)-\frac{\xi}{3}\mathcal{P}\sqrt{x^2-x_0^2}\cos{\theta}A(x)\Big\},\\
        &P_{T_2}=\mathcal{P}\sin{\theta}\cdot F_{T_2}(x)\ /\ \Big\{F(x)-\frac{\xi}{3}\mathcal{P}\sqrt{x^2-x_0^2}\cos{\theta}A(x)\Big\},\\
        &P_{L}=\frac{-F_{IP}(x)+\mathcal{P}\cos{\theta}\cdot F_{AP}(x)}{F(x)-\frac{\xi}{3}\mathcal{P}\sqrt{x^2-x_0^2}\cos{\theta}A(x)},
    \end{split}
\end{equation}
where
\begin{equation}
    \begin{split}
        &F(x)=x(1-x)+\frac{2}{9}\rho\left(4x^2-3x-x_0^2\right)+\eta x_0(1-x),\\
        &A(x)=1-x+\frac{2}{3}\delta\left(4x-4+\sqrt{1-x_0^2}\right),\\
        &F_{T_1}(x)=\frac{1}{12}\bigg[-2\Bigg(\xi^{''}+12\Big(\rho-\frac{3}{4}\Big)\Bigg)(1-x)x_0-3\eta(x^2-x_0^2)+\eta^{''}(-3x^2+4x-x_0^2)\bigg],\\
        &F_{T_2}(x)=\frac{1}{3}\sqrt{x^2-x_0^2}\bigg[3\frac{\alpha^{'}}{\mathcal{A}}(1-x)+2\frac{\beta^{'}}{\mathcal{A}}\sqrt{1-x_0^2}\bigg],\\
        &F_{IP}(x)=\frac{1}{54}\sqrt{x^2-x_0^2}\bigg[9\xi^{'}\Big(-2x+2+\sqrt{1-x_0^2}\Big)+4\xi\Big(\delta-\frac{3}{4}\Big)\Big(4x-4+\sqrt{1-x_0^2}\Big) \bigg],\\
        &F_{AP}(x)=\frac{1}{6}\bigg[\xi^{''}(2x^2-x-x_0^2)+4\Big(\rho-\frac{3}{4}\Big)(4x^2-3x-x_0^2)+2\eta^{''}(1-x)x_0\bigg],
    \end{split}
\end{equation}
are written in terms of the Michel parameters $\rho,\eta,\delta,\xi,\eta^{''},\xi^{'},\xi^{''},\alpha^{'},\beta^{'}$, which are  bilinear combinations of the $g_{\epsilon\omega}^n$ couplings \cite{1,15,19,21,23}.
In the SM, $\rho=\delta=3/4$, $\eta=\eta^{''}=\alpha^{'}=\beta^{'}=0$ and $\xi=\xi^{'}=\xi^{''}=1$.\\
We can obtain the total decay rate integrating over all energy and angular configurations, leading to
\begin{equation}
\label{eqn:tdrr}
    \Gamma_{\ell\rightarrow \ell^{'}}=\frac{\Hat{G}_{\ell\ell^{'}}^2m_\ell^5}{192\pi^3}f(m_{\ell^{'}}^2/m_\ell^2)\left(1+\delta_{RC}^{\ell\ell^{'}}\right),
\end{equation}
where
\begin{equation}
\label{eqn:lowenergyparameter}
    \Hat{G}_{\ell\ell^{'}}\equiv G_{\ell\ell^{'}}\sqrt{1+4\eta\frac{m_{\ell^{'}}}{m_\ell}\frac{g(m_{\ell^{'}}^2/m_\ell^2)}{f(m_{\ell^{'}}^2/m_\ell^2)}}\,,
\end{equation}
 in which  $f(x)=1-8x+8x^3-x^4-12x^2\log(x)$, $g(x)=1+9x-9x^2-x^3+6x(1+x)\log (x)$ and the SM radiative correction $\delta_{RC}^{\ell\ell^{'}}$ has been included \cite{24,25,26}.\\
The normalization $G_{e\mu}$ corresponds to the Fermi coupling $G_F$, measured in the $\mu$ decay.\\
Unlike all other Michel parameters, the dependence of eq.(\ref{eqn:tdrr}) on $\eta$ is kept although it is suppressed by a factor of $(m_{\ell^{'}}/m_\ell)$. Thus, it may be a negligible contribution in the case of $\mu^-\rightarrow e^- \nu_\mu \Bar{\nu_e}$ or $\tau^-\rightarrow e^- \nu_\tau \Bar{\nu_e}$ but not for $\tau^-\rightarrow \mu^- \nu_\tau \Bar{\nu_\mu}$, where it can make measurable distortions of the partial decay width, see e.g.\cite{27}. It is this difference that allows one to determine $\eta$ by comparing the branching ratios of the two $\tau$ leptonic decays. \\
If $e/\mu$ universality is assumed\footnote{Assuming lepton universality leads to a more restrictive value for $\eta$. }, the leptonic decay ratio $B_\mu /B_e$ implies:
\begin{equation}
    \eta=0.016\pm 0.013.
\end{equation}
A non-zero value of $\eta$ would show that there are at least two different couplings with opposite chiralities for the charged leptons. Assuming the V-A coupling $g_{LL}^V$ to be dominant (as supported by data), then the term $\frac{1}{2}\operatorname{Re}[g_{LL}^Vg_{RR}^{S*}]$ is the only one linear in non-standard couplings in the whole spectrum, thus the measurement of $\eta$ is of particular interest for the determination of new couplings, specially the scalar coupling $g_{RR}^{S}$, that usually appears in many extensions of the standard model.\\
The experimental status on the $\tau$-decay Michel parameters \cite{28,29,30,31,32,33,34} together with the more accurate values measured in $\mu$ decay \cite{35,36,37,38} are shown in table \ref{tab:mich}.
\begin{table}[]
\centering
\caption{ Michel parameters and their most accurate determinations.}
\label{tab:mich}
\begin{tabular}{|c|c|c|c|}
\hline
   &$\mu^-\rightarrow e^- \nu_\mu \Bar{\nu_e}$&$\tau^-\rightarrow e^- \nu_\tau \Bar{\nu_e}$&$\tau^-\rightarrow \mu^- \nu_\tau \Bar{\nu_\mu}$\\ [5pt]
\hline
$\rho$& 0.74979 $\pm$  0.00026&0.747 $\pm$ 0.010 & 0.763 $\pm$ 0.020 \\ [10pt]
$\eta$& 0.057 $\pm$ 0.034& ---&0.094 $\pm$  0.073 \\ [10pt]
$\xi$& $1.0009^{+0.0016}_{-0.0007}$&0.994 $\pm$ 0.040 & 1.030 $\pm$ 0.059 \\ [10pt]
$\xi\delta$&$0.7511^{+0.0012}_{-0.0006}$ &0.734 $\pm$ 0.028 & 0.778 $\pm$ 0.037 \\ [10pt]
$\xi^{'}$& 1.00 $\pm$ 0.04&--- &--- \\ [10pt]
$\xi^{''}$& 0.65 $\pm$ 0.36&--- & ---\\ [10pt]
\hline
\end{tabular}
\end{table}
The polarization of the charged lepton emitted in the leptonic $\tau$ decay, related to $\xi^{'}$ and $\xi^{''}$, has never been measured directly. This has be done measuring the $\bar{\eta}$ and $\xi\kappa$ parameters\footnote{Where $\xi^{'}=-\xi-4\xi\kappa+8\xi\delta/3$ and $\xi^{''}=16\rho/3-4\bar{\eta}-3$. } in the radiative leptonic $\tau$-decay \cite{radtaudecay}, since the distribution of the photons emitted by the daughter lepton is sensitive to the lepton polarization \cite{stahl}. Nevertheless the experimental precision does not provide a significant impact on the knowledge about the couplings, yet.\\
In order to established bounds on the couplings, it is convenient to introduce the probabilities \cite{20}
\begin{equation}
    Q_{\epsilon\omega}=\frac{1}{4}|g_{\epsilon\omega}^S|^2+|g_{\epsilon\omega}^V|^2+3(1-\delta_{\epsilon\omega})|g_{\epsilon\omega}^T|^2
\end{equation}
for the decay of an $\omega$-handed $\ell^-$ into an $\epsilon$-handed daughter lepton.\\
The probabilities $ Q_{\epsilon\omega}$ can be extracted from the measurable shape parameters, as follows:
\begin{equation}
    \begin{split}
        &Q_{LL}= \beta^-=\frac{1}{4}\left(-3+\frac{16}{3}\rho-\frac{1}{3}\xi+\frac{16}{9}\xi\delta+\xi^{'}+\xi^{''}\right)\,,\\
        &Q_{RR}=\beta^+ =\frac{1}{4}\left(-3+\frac{16}{3}\rho+\frac{1}{3}\xi-\frac{16}{9}\xi\delta-\xi^{'}+\xi^{''}\right)\,,\\
        &Q_{LR}= \alpha^- + \gamma^-=\frac{1}{4}\left(5-\frac{16}{3}\rho+\frac{1}{3}\xi-\frac{16}{9}\xi\delta+\xi^{'}-\xi^{''}\right)\,,\\
        &Q_{RL}=\alpha^+ + \gamma^+ =\frac{1}{4}\left(5-\frac{16}{3}\rho-\frac{1}{3}\xi+\frac{16}{9}\xi\delta-\xi^{'}-\xi^{''}\right)\,.
    \end{split}
\end{equation}
Upper bounds on any of these probabilities translate into corresponding limits for all couplings with
the given chiralities. The present 90\% CL bounds on the $\mu$-decay couplings and 95\% CL bounds on the $\tau$-decay couplings can be found in \cite{15}.\\
Finally, many processes such as radiative muon and tau decays as well as five-body leptonic decays of muons and tau leptons can be also used to obtain independent constraints and get an additional information about the structure of weak interactions, see e.g.\cite{39,40,41}.
\section{Lepton Decays with Dirac and Majorana Neutrinos}
\label{sec:3}
In this section we analyse the decay amplitude coming from the most general four-lepton interaction Hamiltonian, including the effects due to the Dirac or Majorana nature of neutrinos, that could be measured in the most recent and future experiments.\\
All these studies are based on scenarios where the new sterile neutrinos have non-negligible mixings and some of them require masses low enough to be produced on-shell. Finally, we compare these results and discuss their properties and fundamental differences. Dirac and Majorana distributions are compared in the next section,  \ref{sec:4}.\\
In this framework, the charged weak current interaction is written in the mass eigenstates basis of the charged leptons $\ell$ and the neutrinos $N_j$, after diagonalizing the charged lepton and neutrino mass matrices.\\
Working in the basis where the mass matrix of charged leptons is already diagonalized, the current neutrino ($\nu_{L,R}$) is assumed to be the superposition of the mass-eigenstate neutrinos ($N_j$) with the mass $m_j$, that is,
\begin{equation}
    \nu_{\ell L}=\sum_j U_{\ell j}N_{jL},\quad \nu_{\ell R}=\sum_j V_{\ell j}N_{jR},
\end{equation}
where $j=\{1,2,...,n\}$,  with $n$ the number of  mass-eigenstate neutrinos.\\
As shown by Langacker and London \cite{42}, explicit lepton-number nonconservation still leads to a matrix element equivalent to the one derived from eq.(\ref{eqn:H}).

\subsection{The Effective Decay Rate for Dirac Neutrinos}
\label{subsec:3.1}
In the mass basis, the effective Hamiltonian for the $\ell^{-}\longrightarrow \ell^{'^{-}}\overline{N}_j N_k$ process is written as:
\begin{equation}
\label{eqn:MassHamil1}
\begin{split}
     \mathcal{H}&=4\frac{G_{\ell\ell^{'}}}{\sqrt{2}}\sum_{j,k}\bigg\{ g_{LL}^S\left[\Bar{\ell}^{'}_L V_{\ell^{'}j}N_{jR}\right]\Big[\overline{N}_{kR}V_{\ell k}^{*} \ell_L\Big]+g_{LL}^V\left[\Bar{\ell}^{'}_L \gamma^\mu U_{\ell^{'}j}N_{jL}\right]\Big[\overline{N}_{kL}U_{\ell k}^{*}\gamma_\mu \ell_L\Big]\\
     &+g_{RR}^S\left[\Bar{\ell}^{'}_R U_{\ell^{'}j}N_{jL}\right]\Big[\overline{N}_{kL}U_{\ell k}^{*} \ell_R\Big]
     +g_{RR}^V\left[\Bar{\ell}^{'}_R \gamma^\mu V_{\ell^{'}j}N_{jR}\right]\Big[\overline{N}_{kR}V_{\ell k}^{*}\gamma_\mu \ell_R\Big]\\
     &+g_{LR}^S\left[\Bar{\ell}^{'}_L V_{\ell^{'}j}N_{jR}\right]\Big[\overline{N}_{kL}U_{\ell k}^{*} \ell_R\Big]+g_{LR}^V\left[\Bar{\ell}^{'}_L \gamma^\mu U_{\ell^{'}j}N_{jL}\right]\Big[\overline{N}_{kR}V_{\ell k}^{*}\gamma_\mu \ell_R\Big]\\
     &+g_{LR}^T\left[\Bar{\ell}^{'}_L \frac{\sigma^{\mu\nu}}{\sqrt{2}} V_{\ell^{'}j}N_{jR}\right]\Big[\overline{N}_{kL}U_{\ell k}^{*}\frac{\sigma_{\mu\nu}}{\sqrt{2}} \ell_R\Big]+g_{RL}^S\left[\Bar{\ell}^{'}_R U_{\ell^{'}j}N_{jL}\right]\Big[\overline{N}_{kR}V_{\ell k}^{*} \ell_L\Big]\\
     &+g_{RL}^V\left[\Bar{\ell}^{'}_R \gamma^\mu V_{\ell^{'}j}N_{jR}\right]\Big[\overline{N}_{kL}U_{\ell k}^{*}\gamma_\mu \ell_L\Big]+ g_{RL}^T\left[\Bar{\ell}^{'}_R \frac{\sigma^{\mu\nu}}{\sqrt{2}} U_{\ell^{'}j}N_{jL}\right]\Big[\overline{N}_{kR}V_{\ell k}^{*}\frac{\sigma_{\mu\nu}}{\sqrt{2}} \ell_L\Big]\bigg\}.
     \end{split}
\end{equation}
Note that $\overline{N}$ represents an antineutrino for the Dirac neutrino case, but should be identified with $N$ for the Majorana neutrino case ($N$=$N^c$=$C\overline{N}^T$). Unlike the Majorana case (see section \ref{subsec:3.2}), we will have only one possible first-order Feynman diagram.\\
The differential decay rate for the process $\ell^{-}\longrightarrow \ell^{'^{-}} \overline{N}_j N_k$ is
\begin{equation}
    d\Gamma=\sum_{j,k} d\Gamma^{jk}.
\end{equation}
Explicitly:
\begin{equation}
    d\Gamma=\sum_{j,k} \frac{(2\pi)^4\delta^4(p_1-p_2-p_3-p_4)}{2m_\ell}\frac{d^3p_2 d^3p_3 d^3p_4 }{(2\pi)^32E_2 (2\pi)^32E_3 (2\pi)^32E_4}|\mathcal{M}^D_{jk}|^2,
\end{equation}
where we are taking into account all the possible neutrino mass final states and the sum extends over all energetically allowed neutrino pairs.\\ 
Assuming that the neutrinos are not detected, we can integrate over their momenta. Due to the explicit dependence of the amplitude in the neutrinos momenta, we will have to compute three kinds of phase space integrals, of the form:
\begin{equation}
\label{eqn:tensorintegrals}
\begin{split}
   & I_{\mu\nu}\equiv\int \frac{d^3p_2 d^3p_3}{E_2 E_3}\delta^4(p_2 + p_3 - q)p_{2\mu} p_{3\nu},\\
   & I_{\mu}\equiv\int \frac{d^3p_2 d^3p_3}{E_2 E_3}\delta^4(p_2+p_3-q)p_{(2,3)\mu},\\
   & I\equiv\int \frac{d^3p_2 d^3p_3}{E_2 E_3}\delta^4(p_2 + p_3 - q),
\end{split}
\end{equation}
with $q\equiv p_1-p_4$.\\
We can use the covariance properties of these tensor integrals and the well-known three-body phase space results to compute them, leading to:
\begin{equation}
\begin{split}
        &I_{\mu\nu}=\frac{\pi}{6}[q^2g_{\mu\nu}+2q_\mu q_\nu],\\
        &I_{\mu}=\pi q_\mu,\\
        &I= 2\pi.
\end{split}
\end{equation}
It is important to emphasize that in the phase space integration we are neglecting terms with neutrino mass dependence, in good agreement with the known final lepton energy distribution. This is safe now, because the dependence on neutrino mass effects is quadratic in the phase space calculation, actually of the form $r^2_{jk}=\frac{(m_j+ m_k)^2}{2m_\ell \omega}$; a contribution that will be very small, either by the masses of light neutrinos or by the suppression in the mixing of heavy neutrinos. Furthermore, these effects will have some consequences only at the very end of the energy spectrum, where the statistics is lower. A complete discussion of these effects is given in \cite{12,43,Giannini:2022ilc}. If more precision is needed, the calculation must take into account non-negligible effects due to neutrino masses. For this work, we are dealing with neutrino masses effects up to the matrix element calculation.\\
With this consideration, the final differential decay rate for Dirac neutrinos is given in section \ref{subsec:3.3}.

\subsection{The Effective Decay Rate for Majorana Neutrinos}
\label{subsec:3.2}
\begin{figure}[tbp]
\centering 
\includegraphics[width=.45\textwidth]{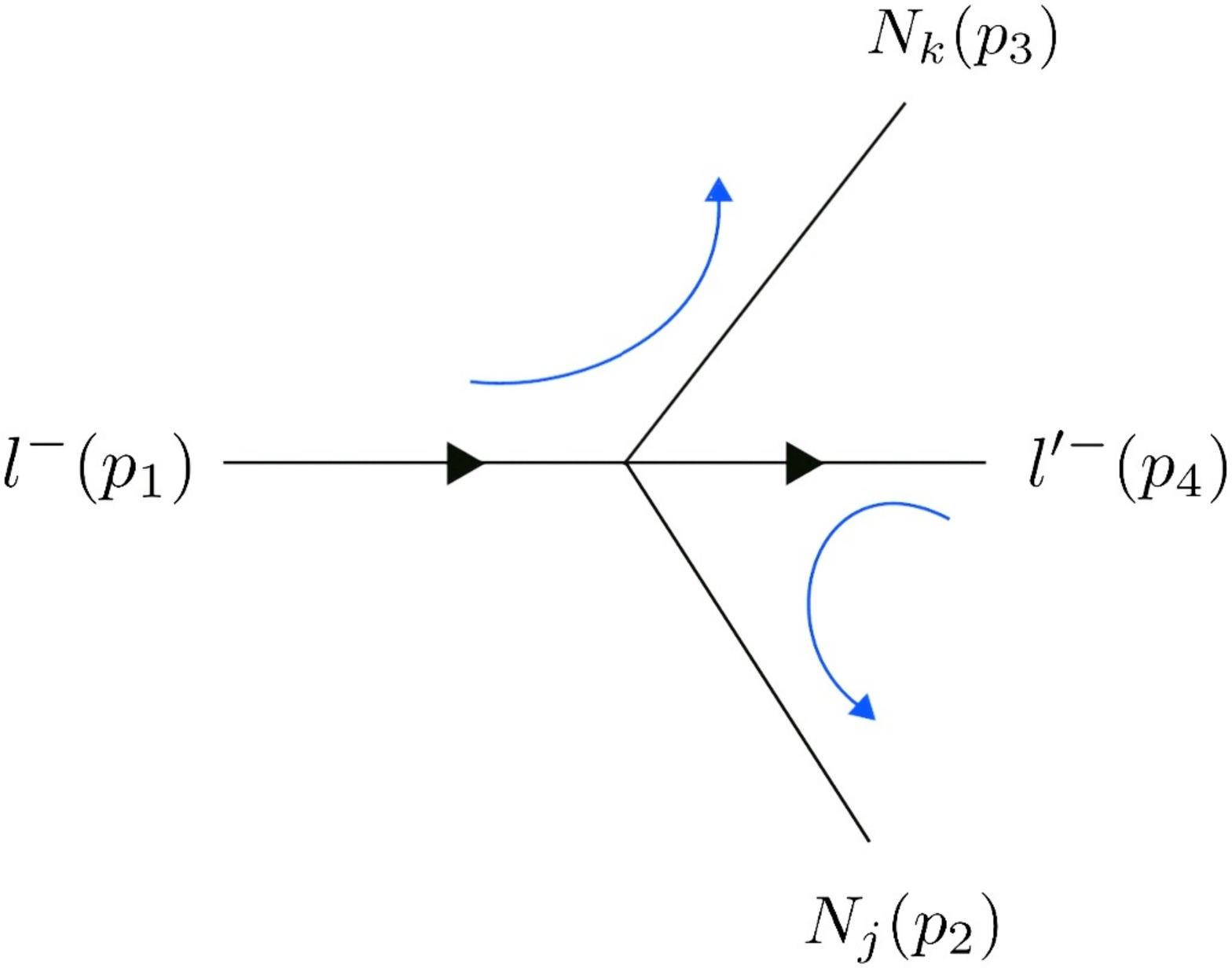}
\hfill
\includegraphics[width=.45\textwidth,origin=c,angle=0]{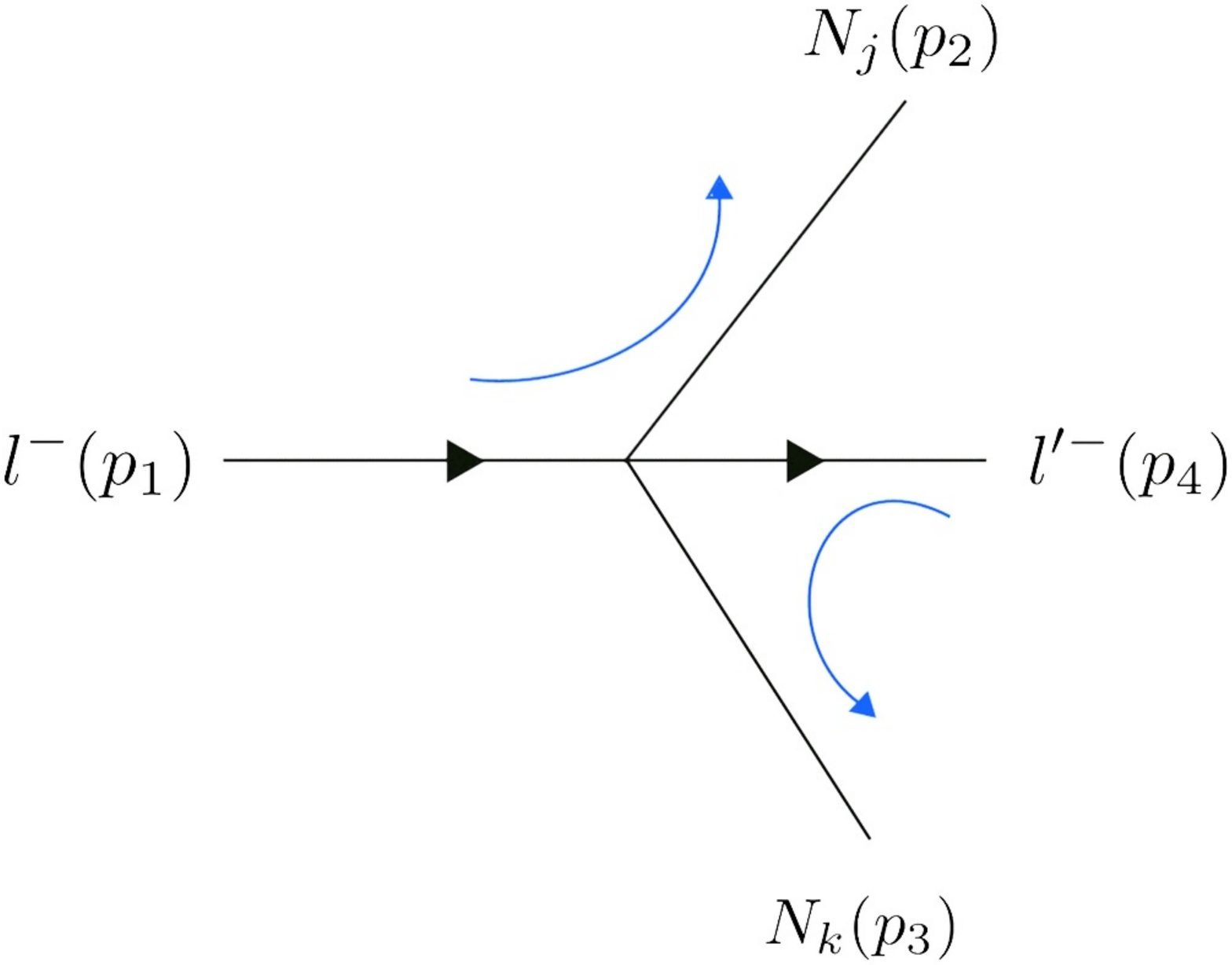}
\caption{\label{fig:1}Feynman diagrams for the $\ell^{-}\longrightarrow \ell^{'^{-}} N_j N_k$ decay. The orientation chosen for the fermion chains is defined by the blue arrows.}
\end{figure}

The Hamiltonian for the case of Majorana neutrinos is
\begin{equation}
\label{eqn:MassHamil2}
\begin{split}
     \mathcal{H}&=4\frac{G_{\ell\ell^{'}}}{\sqrt{2}}\sum_{j,k}\bigg\{ g_{LL}^S V_{\ell^{'}j}V_{\ell k}^{*}\left[\Bar{\ell}^{'}_L N_{jR}\right]\Big[N_{kR} \ell_L\Big]+g_{LL}^V U_{\ell^{'}j} U_{\ell k}^{*}\left[\Bar{\ell}^{'}_L \gamma^\mu N_{jL}\right]\Big[N_{kL}\gamma_\mu \ell_L\Big]\\
     &+g_{RR}^S U_{\ell^{'}j} U_{\ell k}^{*}\left[\Bar{\ell}^{'}_R N_{jL}\right]\Big[N_{kL} \ell_R\Big]
     +g_{RR}^V V_{\ell^{'}j} V_{\ell k}^{*}\left[\Bar{\ell}^{'}_R \gamma^\mu N_{jR}\right]\Big[N_{kR}\gamma_\mu \ell_R\Big]\\
     &+g_{LR}^S V_{\ell^{'}j} U_{\ell k}^{*}\left[\Bar{\ell}^{'}_L N_{jR}\right]\Big[N_{kL} \ell_R\Big]+g_{LR}^V U_{\ell^{'}j} V_{\ell k}^{*}\left[\Bar{\ell}^{'}_L \gamma^\mu N_{jL}\right]\Big[N_{kR}\gamma_\mu \ell_R\Big]\\
     &+g_{LR}^T V_{\ell^{'}j} U_{\ell k}^{*}\left[\Bar{\ell}^{'}_L \frac{\sigma^{\mu\nu}}{\sqrt{2}} N_{jR}\right]\Big[N_{kL}\frac{\sigma_{\mu\nu}}{\sqrt{2}} \ell_R\Big]+g_{RL}^S U_{\ell^{'}j} V_{\ell k}^{*}\left[\Bar{\ell}^{'}_R N_{jL}\right]\Big[N_{kR} \ell_L\Big]\\
     &+g_{RL}^V V_{\ell^{'}j} U_{\ell k}^{*}\left[\Bar{\ell}^{'}_R \gamma^\mu N_{jR}\right]\Big[N_{kL}\gamma_\mu \ell_L\Big]+ g_{RL}^T U_{\ell^{'}j} V_{\ell k}^{*}\left[\Bar{\ell}^{'}_R \frac{\sigma^{\mu\nu}}{\sqrt{2}} N_{jL}\right]\Big[N_{kR}\frac{\sigma_{\mu\nu}}{\sqrt{2}} \ell_L\Big]\bigg\}
     \end{split}
\end{equation}
 Unlike the Dirac case, the properties of Majorana neutrinos have strong consequences in the amplitude. As it has been pointed out in well-known previous works (see the
discussion just before note added of \cite{51}) for Majorana neutrinos the decay
modes $\ell^{-}\longrightarrow \ell^{'^{-}} \overline{N}_j N_k$ and $\ell^{-}\longrightarrow \ell^{'^{-}} \overline{N}_k N_j$ yield the same final states for $j\neq k$ as
well as for $j=k$ (since $\overline{N_i}=N_i$), and hence the amplitudes must be added coherently. Then the possible first order Feynman diagrams for the $\ell^{-}\longrightarrow \ell^{'^{-}} N_j N_k$ decay are shown in figure \ref{fig:1}.\\
 The first diagram leads to the same matrix element as the Dirac case, while the second diagram is only possible in the Majorana neutrino case and we already defined the orientation for each fermion chain (blue arrows).\\
 We define this fermion chain orientation so we can apply the spinors completeness relation directly in the matrix element calculation. This choice will, of course, not  affect the observables, as long as we follow the corresponding Feynman rules for Majorana fermions \cite{44}.\\
 Since the differential decay rate for the process $\ell^{-}\longrightarrow \ell^{'^{-}} N_j N_k$ is
\begin{equation}
    d\Gamma=\frac{1}{2}\sum_{j,k} d\Gamma^{jk}.
\end{equation}
Explicitly:
\begin{equation}
    d\Gamma=\frac{1}{2}\sum_{j,k}\frac{(2\pi)^4\delta^4(p_1-p_2-p_3-p_4)}{2m_\ell}\frac{d^3p_2 d^3p_3 d^3p_4 }{(2\pi)^32E_2 (2\pi)^32E_3 (2\pi)^32E_4}|\mathcal{M}_{jk}|^2,
\end{equation}
where the 1/2 factor that appears in the Majorana case has two different origins. For the $j=k $ case it is a statistical factor due to the presence of indistinguishable
fermions in the final state, while for the $j\neq k$ case it arrives because of double counting, since the sum $\sum\limits_{j,k}$ is not restricted to $j\leq k$. The same discussion can be found in section 11.2.2 of \cite{52}.\\ 
Then, after integrating over the neutrinos momenta, the decay rate will have the following dependence on the amplitude:\footnote{Where $\mathcal{M}_{jk}^D$ and $\mathcal{M}_{jk}^M$ are related by the exchange $j\leftrightarrow k$ once the neutrinos momenta are integrated out.}
\begin{equation}
\begin{split}
   d\Gamma&\propto \frac{1}{2}\sum_{j,k} |\mathcal{M}_{jk}^D-\mathcal{M}_{jk}^M|^2\\
   & =\frac{1}{2}\sum_{j,k}\Big\{ |\mathcal{M}_{jk}^D|^2+|\mathcal{M}_{jk}^M|^2-2\operatorname{Re}(\mathcal{M}_{jk}^D \mathcal{M}_{jk}^{M*})\big\}\\
   &=\sum_{j,k} |\mathcal{M}_{jk}^D|^2-\sum_{j,k}\operatorname{Re}(\mathcal{M}_{jk}^D \mathcal{M}_{jk}^{M*}).
\end{split}
\end{equation}
Thus the fundamental difference between Dirac and Majorana cases is precisely the last interference term, which is sometimes called the $Majorana$ $term$ \cite{52}. The relative sign between $\mathcal{M}_{jk}^D$ and $\mathcal{M}_{jk}^M$ stems from the application of Wick's theorem in presence of Majorana fermions (see, e.g. Ref.~\cite{44}).\\
This fact allows us to write the final differential decay rate for both cases in a single expression, where Dirac and Majorana nature is distinguishable by the Majorana term, that can be added with the implementation of a flag parameter $\epsilon=0,1$. 

\clearpage
\subsection{Final Distribution (Dirac and Majorana)}
\label{subsec:3.3}
In order to be more efficient, we write our final result in the PDG parametrization, which is currently the most useful. Also we distinguish the Majorana contribution, by explicit calculation of the Majorana term, with the help of a flag parameter $\epsilon=0,1$. Finally we rename dummy indices wherever possible in order to make our expressions clearer and easier to use for general applications.\\
The differential decay probability to obtain a final charged lepton with (reduced) energy between $x$ and $x+dx$, emitted in the direction $\hat{z}$ at an angle between $\theta$ and $\theta+d\theta$ with respect to the initial lepton polarization vector $\mathcal{P}$, and with its spin parallel to the arbitrary direction $\hat{\zeta}$, neglecting radiative corrections and taking into account finite Dirac ($\epsilon=0$) or Majorana ($\epsilon=1$) neutrino masses is given by
\begin{equation}
    \begin{split}
    \label{eqn:ResultadoFinalMajoranayDirac}
        \frac{d\Gamma}{dx d\cos{\theta}}&=\sum_{j,k}\frac{m_\ell}{4 \pi^3} \omega^4 G_{\ell\ell^{'}}^2\sqrt{x^2-x_0^2}\\
        &\times\Big(\left(F_{IS}(x)+F_{IS}'(x)+F_{IS}''(x)\right)-\mathcal{P}\cos{\theta}\big(F_{AS}(x)+F_{AS}'(x)+F_{AS}''(x)\big)\Big)\\
        &\times\big[1+\hat{\zeta}\cdot \Vec{\mathcal{P}}_{\ell^{'}}(x,\theta)\big],
    \end{split}
\end{equation}
where we recall the polarization vector $\Vec{\mathcal{P}}_{\ell^{'}}$ is
\begin{equation}
    \Vec{\mathcal{P}}_{\ell^{'}}=P_{T_1}\cdot\hat{x}+P_{T_2}\cdot\hat{y}+P_{L}\cdot\hat{z}.
\end{equation}
We also remind that here $\hat{x},\hat{y}$ and $\hat{z}$ are unit vectors defined as:
\begin{equation}
    \begin{split}
        &\hat{z}\quad \text{is along the $\ell^{'}$momentum $\Vec{p}_{\ell^{'}}$,}\\
        \frac{\hat{z}\times\Vec{\mathcal{P}}_{\ell^{'}}}{|\hat{z}\times\Vec{\mathcal{P}}_{\ell^{'}}|}= &\hat{y}\quad \text{is transverse to $\Vec{p}_{\ell^{'}}$ and perpendicular to the decay plane,}\\
        \hat{y}\times\hat{z}=& \hat{x}\quad \text{is transverse to $\Vec{p}_{\ell^{'}}$ and in the decay plane,}
    \end{split}
\end{equation}
and the components of $\Vec{\mathcal{P}}_{\ell^{'}}$ are, respectively,
\begin{equation}
    \begin{split}
        &P_{T_1}=\mathcal{P}\sin{\theta}\cdot \big( F_{T_1}(x)+F_{T_1}'(x)+F_{T_1}''(x)\big)\ /\ N,\\
        &P_{T_2}=\mathcal{P}\sin{\theta}\cdot \big( F_{T_2}(x)+F_{T_2}'(x)+F_{T_2}''(x)\big)\ /\ N,\\
        &P_{L}=\Big(-\big(F_{IP}(x)+F_{IP}'(x)+F_{IP}''(x) \big)+\mathcal{P}\cos{\theta}\cdot \big(F_{AP}(x)+F_{AP}'(x)+F_{AP}''(x)\big)\Big)\ /\ N\,,
    \end{split}
\end{equation}
with $N$ the normalization factor $N=\big(F_{IS}(x)+F_{IS}'(x)+F_{IS}''(x)\big)-\mathcal{P}\cos{\theta}\big(F_{AS}(x)+F_{AS}'(x)+F_{AS}''(x)\big)$.\\
The corresponding functions are defined as:
\begin{equation}
    \begin{split}
     &F_{IS}(x)=x(1-x)(I)_{jk}+\frac{2}{9}(\rho)_{jk}\left(4x^2-3x-x_0^2\right)+(\eta)_{jk} x_0(1-x),\\
        &F_{AS}(x)=\frac{(\xi)_{jk}}{3}\sqrt{x^2-x_0^2}\left(1-x+\frac{2}{3}(\delta)_{jk}\left(4x-4+\sqrt{1-x_0^2}\right)\right),\\
        &F_{IS}'(x)=\frac{1}{4}\frac{m_j}{m_\ell}\operatorname{Re}\left[(\kappa_L^+)_{jk} \left(\color{black} x\left(1+\sqrt{1-x_0^2}\right)-x_0^2\right)\color{black}-(\kappa_R^+)_{kj}\left(\color{black}x_0(1-x)+x_0\sqrt{1-x_0^2}\right)\color{black}\right],\\
        &F_{AS}'(x)=\frac{1}{4}\frac{m_j}{m_\ell}\sqrt{x^2-x_0^2}\operatorname{Re}\left[(\kappa_R^-)_{kj} x_0+(\lambda_L^-)_{jk}\left(1+\sqrt{1-x_0^2}\right) \right],\\
        &F_{IS}''(x)=-\frac{1}{2}\frac{m_j m_k}{m_\ell^2}\left(1+\sqrt{1-x_0^2}\right)\operatorname{Re}\left(x (C^{+})_{jk}+x_0 (H^{+})_{jk}\right),\\
        &F_{AS}''(x)=\frac{1}{2}\frac{m_j m_k}{m_\ell^2}\left(1+\sqrt{1-x_0^2}\right)\sqrt{x^2-x_0^2} \operatorname{Re}(C^{-})_{jk},\\
     &F_{T_1}(x)=\frac{1}{12}\bigg[-2\left((\xi^{''})_{jk}+12\left((\rho)_{jk}-\frac{3}{4}(I)_{jk}\right)\right)(1-x)x_0-3(\eta)_{jk}\left(x^2-x_0^2\right)+(\eta^{''})_{jk}\\
     &\quad\quad\quad\quad\left(-3x^2+4x-x_0^2\right)\bigg],\\
        &F_{T_2}(x)=\frac{1}{3}\sqrt{x^2-x_0^2}\left[3\left(\frac{\alpha^{'}}{\mathcal{A}}\right)_{jk}(1-x)+2\left(\frac{\beta^{'}}{\mathcal{A}}\right)_{jk}\sqrt{1-x_0^2}\right],\\
        &F_{IP}(x)=\frac{1}{54}\sqrt{x^2-x_0^2}\left[9(\xi^{'})_{jk}\left(-2x+2+\sqrt{1-x_0^2}\right)+4(\xi)_{jk}\left((\delta)_{jk}-\frac{3}{4}\right)\left(4x-4+\sqrt{1-x_0^2}\right) \right],\\
        &F_{AP}(x)=\frac{1}{6}\left[(\xi^{''})_{jk}\left(2x^2-x-x_0^2\right)+4\left((\rho)_{jk}-\frac{3}{4}(I)_{jk}\right)\left(4x^2-3x-x_0^2\right)+2(\eta^{''})_{jk}(1-x)x_0\right],\\
        &F_{T_1}'(x)=\frac{1}{4}\frac{m_j}{m_\ell}\operatorname{Re}\left[(\lambda_L^+)_{jk}\left(x_0(1-x)+x_0\sqrt{1-x_0^2}\right)- (\lambda_R^+)_{kj}\left(x\left(1+\sqrt{1-x_0^2}\right)-x_0^2 \right)\right],\\
        &F_{T_2}'(x)=\frac{1}{4}\frac{m_j}{m_\ell}\sqrt{x^2-x_0^2}\operatorname{Im}\left[(\lambda_R^-)_{kj}\left(1+\sqrt{1-x_0^2}\right)+(\lambda_L^-)_{jk}x_0\right],\\
        &F_{IP}'(x)=\frac{1}{4}\frac{m_j}{m_\ell}\sqrt{x^2-x_0^2}\operatorname{Re}\left[(\lambda_R^-)_{kj}x_0+(\kappa_L^-)_{jk}\left(1+\sqrt{1-x_0^2}\right)\right],\\
        &F_{AP}'(x)=\frac{1}{4}\frac{m_j}{m_\ell}\operatorname{Re}\left[(\lambda_L^+)_{jk}\left(x\left(1+\sqrt{1-x_0^2}\right)-x_0^2\right)-(\lambda_R^+)_{kj}\left(x_0(1-x)+x_0\sqrt{1-x_0^2}\right)\right],\\
         &F_{T_1}''(x)=\frac{1}{2}\frac{m_j m_k}{m_\ell^2}\left(1+\sqrt{1-x_0^2}\right)\operatorname{Re}\left(x_0 (C^{'+})_{jk}-2x(J^{+})_{jk} \right),\\
        &F_{T_2}''(x)=-\frac{m_j m_k}{m_\ell^2}\left(1+\sqrt{1-x_0^2}\right)\operatorname{Im}(J^{+})_{jk}\sqrt{x^2-x_0^2},\\
        &F_{IP}''(x)=-\frac{1}{2}\frac{m_j m_k}{m_\ell^2}\left(1+\sqrt{1-x_0^2}\right)\operatorname{Re}(C^{'-})_{jk}\sqrt{x^2-x_0^2},\\
        &F_{AP}''(x)=\frac{1}{2}\frac{m_j m_k}{m_\ell^2}\left(1+\sqrt{1-x_0^2}\right)\operatorname{Re}\left(x (C^{'+})_{jk}-2x_0 (J^{+})_{jk} \right),
    \end{split}
\end{equation}
where the parameters are  bilinear combinations of the coupling constants $g_{lm}^n$, given by
\begin{equation*}
\begin{split}
&(I)_{jk}=\frac{1}{4}(|(f_{RR}^S)_{jk}|^2+|(f_{RL}^S)_{jk}|^2+|(f_{LR}^S)_{jk}|^2+|(f_{LL}^S)_{jk}|^2)+3(|(f_{LR}^T)_{jk}|^2+|(f_{RL}^T)_{jk}|^2)+(|(f_{RR}^V)_{jk}|^2\\
&\quad\quad\quad+|(f_{RL}^V)_{jk}|^2+|(f_{LR}^V)_{jk}|^2+|(f_{LL}^V)_{jk}|^2)+\frac{\epsilon}{8}\operatorname{Re}\color{blue}\Big[\color{black}12(f_{LR}^{T})_{jk}(f_{LR}^{S})^*_{kj}+12(f_{LR}^{T})_{jk}(f_{LR}^{T})^*_{kj}\\
&\quad\quad\quad+8(f_{RL}^{V})_{jk}(f_{RL}^{V})^*_{kj}-(f_{LR}^{S})_{jk}(f_{LR}^{S})^*_{kj}+8(f_{LL}^{S})_{jk}(f_{LL}^{V})^*_{kj}+(L\leftrightarrow R) \color{blue}\Big]\color{black},\\
&(\rho)_{jk}=\frac{3}{4}(|(f_{LL}^V)_{jk}|^2+\frac{1}{4}|(f_{LL}^S)_{jk}|^2)+\frac{3}{16}|(f_{LR}^S)_{jk}-2(f_{LR}^T)_{jk}|^2+(R \leftrightarrow L)+\epsilon\frac{3}{16}\operatorname{Re}\color{blue}\Big[\color{black} -(f_{LR}^{S})_{jk}\\
&\quad\quad\quad(f_{LR}^{S})^*_{kj}+4(f_{LR}^{S})_{jk}(f_{LR}^{T})^*_{kj}+4(f_{LL}^{S})_{jk}(f_{LL}^{V})^*_{kj}-4(f_{LR}^{T})_{jk}(f_{LR}^{T})^*_{kj}+(L\leftrightarrow R) \color{blue}\Big]\color{black},\\
        &(\xi)_{jk}=3\big(|(f_{LR}^V)_{jk}|^2+\frac{1}{16}|(f_{LR}^S)_{jk}+6(f_{LR}^T)_{jk}|^2\big)+|(f_{LL}^V)_{jk}|^2+\frac{1}{4}|(f_{LL}^S)_{jk}|^2-\frac{7}{16}|(f_{LR}^S)_{jk}-2(f_{LR}^T)_{jk}|^2\\
        &\quad\quad\quad-(R \leftrightarrow L)+\epsilon\operatorname{Re}\color{blue}\Big[\color{black}-(f_{RR}^{S})_{jk}(f_{RR}^{V})^*_{kj}+\frac{17}{2}(f_{LR}^{T})_{jk}(f_{LR}^{T})^*_{kj}+\frac{1}{2}(f_{LR}^{S})_{jk}(f_{LR}^{T})^*_{kj}+3(f_{LR}^{V})_{jk}\\
&\quad\quad\quad(f_{LR}^{V})^*_{kj}+\frac{5}{8}(f_{LR}^{S})_{jk}(f_{LR}^{S})^*_{kj}-(L\leftrightarrow R) \color{blue}\Big]\color{black},\\
        &(\xi\delta)_{jk}=\frac{3}{4}(|(f_{LL}^V)_{jk}|^2+\frac{1}{4}|(f_{LL}^S)_{jk}|^2)-\frac{3}{16}|(f_{LR}^S)_{jk}-2(f_{LR}^T)_{jk}|^2-(R \leftrightarrow L)+\epsilon\frac{3}{4}\operatorname{Re}\color{blue}\Big[\color{black} -(f_{RR}^{S})_{jk}\\
&\quad\quad\quad(f_{RR}^{V})^*_{kj}+(f_{LR}^{T})_{jk}(f_{LR}^{T})^*_{kj}-(f_{LR}^{S})_{jk}(f_{LR}^{T})^*_{kj}+\frac{1}{4}(f_{LR}^{S})_{jk}(f_{LR}^{S})^*_{kj}-(L\leftrightarrow R)\color{blue}\Big]\color{black},\\
        &(\eta)_{jk}=\frac{1}{2}\operatorname{Re}[(f_{LL}^V)_{jk}(f_{RR}^{S})^*_{jk}+(f_{RR}^V)_{jk}(f_{LL}^{S})^*_{jk}+(f_{LR}^V)_{jk}((f_{RL}^{S})^*_{jk}+6(f_{RL}^{T})^*_{jk})+(f_{RL}^V)_{jk}((f_{LR}^{S})^*_{jk}\\
        &\quad\quad\quad+6(f_{LR}^{T})^*_{jk})]+\frac{\epsilon}{8}\operatorname{Re}\color{blue}\Big[\color{black}4(f_{LR}^{S})_{jk}(f_{RL}^{V})^*_{kj}+24(f_{LR}^{T})_{jk}(f_{RL}^{V})^*_{kj}+(f_{LL}^{S})_{jk}(f_{RR}^{S})^*_{kj}+4(f_{LL}^{V})_{jk}\\
&\quad\quad\quad(f_{RR}^{V})^*_{kj}+(L\leftrightarrow R) \color{blue}\Big]\color{black},\\
         &(\xi^{'})_{jk}=-\frac{1}{4}\big(|(f_{RR}^S)_{jk}|^2+4|(f_{RR}^V)_{jk}|^2-|(f_{LR}^S)_{jk}|^2-4|(f_{LR}^V)_{jk}|^2-12|(f_{LR}^T)_{jk}|^2\big)-(R\leftrightarrow L)\\
         &\quad\quad\quad+\epsilon\operatorname{Re}\color{blue}\Big[\color{black}(f_{LL}^{S})_{jk}(f_{LL}^{V})^*_{kj}+\frac{3}{2}(f_{LR}^{S})_{jk}(f_{LR}^{T})^*_{kj}+\frac{3}{2}(f_{LR}^{T})_{jk}(f_{LR}^{T})^*_{kj}+(f_{LR}^{V})_{jk}(f_{LR}^{V})^*_{kj}\\
        &\quad\quad\quad-\frac{1}{8}(f_{LR}^{S})_{jk}(f_{LR}^{S})^*_{kj}-(L\leftrightarrow R) \color{blue}\Big]\color{black},\\
        &(\xi^{''})_{jk}=\frac{1}{4}\Big(|(f_{RR}^S)_{jk}|^2+4|(f_{RR}^V)_{jk}|^2-|(f_{LR}^S)_{jk}|^2+12|(f_{LR}^V)_{jk}|^2+20|(f_{LR}^T)_{jk}|^2+16\operatorname{Re}((f_{LR}^{S})_{jk}\\
        &\quad\quad\quad(f_{LR}^{T})^*_{jk}) \Big)+(R \leftrightarrow L)+\epsilon\operatorname{Re}\color{blue}\Big[\color{black}\frac{1}{2}(f_{LR}^{S})_{jk}(f_{LR}^{T})^*_{kj}+\frac{17}{2}(f_{LR}^{T})_{jk}(f_{LR}^{T})^*_{kj}+(f_{LL}^{S})_{jk}(f_{LL}^{V})^*_{kj}\\
        &\quad\quad\quad+\frac{5}{8}(f_{RL}^{S})_{jk}(f_{RL}^{S})^*_{kj}+3(f_{RL}^{V})_{jk}(f_{RL}^{V})^*_{kj}+(L\leftrightarrow R) \color{blue}\Big]\color{black},\\
        &(\eta^{''})_{jk}=\frac{1}{2}\operatorname{Re}\big(3(f_{LR}^{V})_{jk}((f_{RL}^{S})^*_{jk}+6(f_{RL}^{T})^*_{jk})-(f_{LL}^{V})_{jk}(f_{RR}^{S})^*_{jk}\big)+(R \leftrightarrow L)+\frac{\epsilon}{2}\operatorname{Re}\color{blue}\Big[\color{black} 3(f_{LR}^{S})_{jk}(f_{RL}^{V})^*_{kj}\\
        &\quad\quad\quad+18(f_{LR}^{T})_{jk}(f_{RL}^{V})^*_{kj}-\frac{1}{4}(f_{LL}^{S})_{jk}(f_{RR}^{S})^*_{kj}-(f_{LL}^{V})_{jk}(f_{RR}^{V})^*_{kj}+(L\leftrightarrow R)\Big]\color{blue}\Big]\color{black},\\
        &\left(\frac{\alpha^{'}}{\mathcal{A}}\right)_{jk}=\frac{1}{2}\operatorname{Im}\big((f_{LR}^{V})_{jk}((f_{RL}^{S})^*_{jk}+6(f_{RL}^{T})^*_{jk})\big)-(R\leftrightarrow L)+\epsilon\operatorname{Im}\color{blue}\Big[\color{black}\frac{1}{2}(f_{RL}^{V})^*_{jk}((f_{LR}^{S})_{kj}+6(f_{LR}^{T})_{kj})\\
        &\quad\quad\quad\quad-(L\leftrightarrow R) \color{blue}\Big]\color{black},\\ &\left(\frac{\beta^{'}}{\mathcal{A}}\right)_{jk}=-\frac{1}{4}\operatorname{Im}\big((f_{LL}^{V})_{jk}(f_{RR}^{S})^*_{jk}\big)-(R\leftrightarrow L)+\epsilon\operatorname{Im}\color{blue}\Big[\color{black}-\frac{1}{2} (f_{LL}^{V})_{jk}(f_{RR}^{V})^*_{kj}-\frac{1}{8}(f_{LL}^{S})_{jk}(f_{RR}^{S})^*_{kj}\color{blue}\Big]\color{black},
\end{split}
\end{equation*}
\begin{equation}
\begin{split}
\label{completepararesults}
    &(\kappa^{\pm}_N)_{jk}=(f_{NN}^{S})_{kj} (f_{LR}^{S})^*_{kj}-2(f_{NN}^{V})_{kj} (f_{LR}^{V})^*_{kj}- (f_{NN}^{S})_{jk} (f_{LR}^{V})^*_{jk}-(f_{NN}^{V})_{jk} ((f_{LR}^{S})^*_{jk}-6(f_{LR}^{T})^*_{jk})\\
        &\quad\quad\quad\pm (R \leftrightarrow L)+\epsilon\color{blue}\Big[\color{black}-2(f_{NN}^{V})_{kj}(f_{LR}^{V})^*_{jk}-\frac{1}{2}(f_{NN}^{S})_{kj}(f_{LR}^{S})^*_{jk}+3(f_{NN}^{S})_{kj}(f_{LR}^{T})^*_{jk}+2(f_{NN}^{V})_{jk}\\
        &\quad\quad\quad(f_{LR}^{S})^*_{kj}-(f_{NN}^{S})_{jk}(f_{LR}^{V})^*_{kj}\pm(L\leftrightarrow R) \color{blue}\Big]\color{black},\\
        &(\lambda^{\pm}_N)_{jk}=-(f_{NN}^{S})_{jk}(f_{LR}^{V})^*_{jk}+(f_{NN}^{V})_{jk}((f_{LR}^{S})^*_{jk}+2(f_{LR}^{T})^*_{jk})+2(f_{NN}^{S})_{kj}(f_{LR}^{T})^*_{kj}-2(f_{NN}^{V})_{kj}(f_{LR}^{V})^*_{kj}\\
        &\quad\quad\quad\pm (R \leftrightarrow L)+\epsilon\color{blue}\Big[\color{black}-2(f_{NN}^{V})_{kj}(f_{LR}^{V})^*_{jk}+\frac{1}{2}(f_{NN}^{S})_{kj}(f_{LR}^{S})^*_{jk}+(f_{NN}^{S})_{kj}(f_{LR}^{T})^*_{jk}+4(f_{NN}^{V})_{jk}\\
        &\quad\quad\quad(f_{LR}^{T})^*_{kj}-(f_{NN}^{S})_{jk}(f_{LR}^{V})^*_{kj}\pm(L \leftrightarrow R) \color{blue}\Big]\color{black},\\
    &(C^{\pm})_{jk}= (f_{LL}^{S})_{jk}(f_{LL}^{V})^*_{jk}+(f_{RL}^{V})^*_{jk}((f_{RL}^{S})_{jk}+6(f_{RL}^{T})_{jk})\pm (R \leftrightarrow L)+\epsilon\color{blue}\Big[\color{black}(f_{LL}^{V})_{jk}(f_{LL}^{V})^*_{kj}+\frac{1}{4}(f_{LL}^{S})_{jk}\\
        &\quad\quad\quad(f_{LL}^{S})^*_{kj}+(f_{RL}^{S})_{jk}(f_{RL}^{V})^*_{kj}+6(f_{RL}^{V})_{jk}(f_{RL}^{T})^*_{kj}\pm(L\leftrightarrow R) \color{blue}\Big]\color{black},\\
     &(C^{'\pm})_{jk}= (f_{LL}^{S})_{jk}(f_{LL}^{V})^*_{jk}-(f_{RL}^{V})^*_{jk}((f_{RL}^{S})_{jk}+6(f_{RL}^{T})_{jk})\pm (R \leftrightarrow L)+\epsilon\color{blue}\Big[\color{black} \frac{1}{4}(f_{LL}^{S})_{jk}(f_{LL}^{S})^*_{kj}\\
    & \quad\quad\quad+(f_{LL}^{V})_{jk}(f_{LL}^{V})^*_{kj}-(f_{RL}^{S})_{jk}(f_{RL}^{V})^*_{kj}-6(f_{RL}^{V})_{jk}(f_{RL}^{T})^*_{kj}\pm(L \leftrightarrow R) \color{blue}\Big]\color{black},\\
    &(J^{+})_{jk}= (f_{LR}^{S})_{jk}(f_{RL}^{T})^*_{jk}+(f_{LR}^{T})_{jk}(f_{RL}^{S})^*_{jk}+2(f_{LR}^{V})_{jk}(f_{RL}^{V})^*_{jk}+4(f_{LR}^{T})_{jk}(f_{RL}^{T})^*_{jk}+\epsilon\color{blue}\Big[\color{black} \frac{1}{4}(f_{LR}^{S})_{kj}\\
    &\quad\quad\quad(f_{RL}^{S})^*_{jk}+\frac{1}{2}(f_{LR}^{S})_{kj}(f_{RL}^{T})^*_{jk}+\frac{1}{2}(f_{LR}^{T})_{kj}(f_{RL}^{S})^*_{jk}+5(f_{LR}^{T})_{kj}(f_{RL}^{T})^*_{jk}+2(f_{LR}^{V})_{jk}(f_{RL}^{V})^*_{kj}\color{blue}\Big]\color{black},\\
    &(H^{+})_{jk}= (f_{LL}^{S})_{jk}(f_{RR}^{S})^*_{jk}+4(f_{LL}^{V})_{jk}(f_{RR}^{V})^*_{jk}+(f_{LR}^{S})_{jk}(f_{RL}^{S})^*_{jk}+12(f_{LR}^{T})_{jk}(f_{RL}^{T})^*_{jk}\\
    &\quad\quad\quad+4(f_{LR}^{V})_{jk}(f_{RL}^{V})^*_{jk}+\epsilon\color{blue}\Big[\color{black}2(f_{LL}^{V})_{jk}(f_{RR}^{S})^*_{kj}-\frac{1}{4}(f_{LR}^{S})_{jk}(f_{RL}^{S})^*_{kj}+3(f_{LR}^{T})_{jk}(f_{RL}^{S})^*_{kj}+3(f_{LR}^{T})_{jk}\\
        &\quad\quad\quad(f_{RL}^{T})^*_{kj}+2(f_{LR}^{V})_{jk}(f_{RL}^{V})^*_{kj}+(L\leftrightarrow R) \color{blue}\Big]\color{black},
\end{split}
\end{equation}
where we defined the constants $(f_{lm}^{n})_{jk}$ as:
\begin{equation}
    \begin{split}
        &(f_{LL}^{S})_{jk}\equiv g_{LL}^{S}V_{\ell^{'}j}V_{\ell k}^{*},\quad\quad\quad (f_{RR}^{S})_{jk}\equiv g_{RR}^{S}U_{\ell^{'}j} U_{\ell k}^{*},\\
        &(f_{LL}^{V})_{jk}\equiv g_{LL}^{V}U_{\ell^{'}j} U_{\ell k}^{*},\quad\quad\quad  (f_{RR}^{V})_{jk}\equiv g_{RR}^{V}V_{\ell^{'}j} V_{\ell k}^{*},\\
        &(f_{LR}^{S})_{jk}\equiv g_{LR}^{S}V_{\ell^{'}j} U_{\ell k}^{*},\quad\quad\quad (f_{RL}^{S})_{jk}\equiv g_{RL}^{S} U_{\ell^{'}j} V_{\ell k}^{*},\\
        &(f_{LR}^{V})_{jk}\equiv g_{LR}^{V}U_{\ell^{'}j} V_{\ell k}^{*},\quad\quad\quad (f_{RL}^{V})_{jk}\equiv g_{RL}^{V}V_{\ell^{'}j} U_{\ell k}^{*},\\
        &(f_{LR}^{T})_{jk}\equiv g_{LR}^{T} V_{\ell^{'}j} U_{\ell k}^{*},\quad\quad\quad (f_{RL}^{T})_{jk}\equiv g_{RL}^{T}U_{\ell^{'}j} V_{\ell k}^{*},
    \end{split}
\end{equation}
together with the normalization:
\enlargethispage{2\baselineskip}
\begin{equation}
\begin{split}
    1&=(I)_{jk}\equiv \frac{\mathcal{A}}{16}.
    \end{split}
\end{equation}
Finally, integrating over all energy and angular configurations we obtained:
\begin{equation}
\begin{split}
    \Gamma_{\ell\rightarrow \ell^{'}}&=\sum_{j,k} \frac{m_\ell w^4}{\pi^3}G_{\ell\ell^{'}}^2\int_{x_0}^1\sqrt{x^2-x_0^2}\Big(F_{IS}(x)+F'_{IS}(x)+F''_{IS}(x)\Big)dx
    \end{split}
\end{equation}
\begin{equation}
    \Gamma_{\ell\rightarrow \ell^{'}}=\sum_{j,k} \frac{\Hat{G}_{\ell\ell^{'}}^2m_\ell^5}{192\pi^3}f(m_{\ell^{'}}^2/m_\ell^2)\left(1+\delta_{RC}^{\ell\ell^{'}}\right),
\end{equation}
where
\begin{equation}
\begin{split}
    \Hat{G}_{\ell\ell^{'}}\equiv& G_{\ell\ell^{'}}\Bigg\{(I)_{jk}+4(\eta)_{jk}\frac{m_{\ell^{'}}}{m_\ell}\frac{g(m_{\ell^{'}}^2/m_\ell^2)}{f(m_{\ell^{'}}^2/m_\ell^2)}-2\frac{m_j}{m_\ell}\left[(\kappa_L^+)_{jk}\frac{f'(m_{\ell^{'}}^2/m_\ell^2)}{f(m_{\ell^{'}}^2/m_\ell^2)}+(\kappa_R^+)_{kj}\frac{m_{\ell^{'}}}{m_\ell}\frac{g'(m_{\ell^{'}}^2/m_\ell^2)}{f(m_{\ell^{'}}^2/m_\ell^2)}\right]\\
    &-4\frac{m_j m_k}{m_\ell^2}\left[(C^+)_{jk}\frac{f''(m_{\ell^{'}}^2/m_\ell^2)}{f(m_{\ell^{'}}^2/m_\ell^2)}+3(H^+)_{jk} \frac{m_{\ell^{'}}}{m_\ell}\frac{g''(m_{\ell^{'}}^2/m_\ell^2)}{f(m_{\ell^{'}}^2/m_\ell^2)} \right]\Bigg\}^{1/2},
    \end{split}
\end{equation}
with the functions defined as:
\begin{equation}
    \begin{split}
    &f(x)=1-8x-12x^2\log(x)+8x^3-x^4,\\
    &f'(x)=-1+6x-2x^3+3x^2\left(4 \ \text{arctanh} \left(\frac{x-1}{x+1}\right)-1\right),\\
    &f''(x)=1-3x+3x^2-x^3,\\
&g(x)=1+9x-9x^2-x^3+6x(1+x)\log (x),\\
&g'(x)=2-6x^2+x^3+3x\left( 4 \ \text{arctanh}\left(\frac{x-1}{x+1}\right)+1\right),\\
&g''(x)=1-x^2+2x\log(x).
    \end{split}
\end{equation}
As we can see, the $(I)_{jk}$ parameter defines the normalization of the $G_{\ell\ell^{'}}$ coupling, while the other parameters entering as  suppressed corrections.\\
Also we have linear and quadratic neutrino mass terms, each of those having the same behavior as the ones obtained in the massless case, i.e., a parameter without suppression and one parameter that is suppressed by the factor $m_{\ell^{'}}/m_\ell$ that is usually named as low-energy parameter; we remind this could be more important in tau decays. Nevertheless, the neutrino mass suppression makes these contributions to the $G_{\ell\ell^{'}}$ coupling almost negligible.

\section{Dirac vs Majorana Distribution}
\label{sec:4}
Looking at the final charged-lepton distribution we can highlight several things. We now have completely new parameters that fully describe the linear ($\kappa^{\pm}_N, \lambda^{\pm}_N$) and quadratic ($C^{\pm}, C'^{\pm}, J^{\pm}$, $H^{\pm}$) neutrino mass dependence of the differential decay rate. 
The parameters superscript ($\pm$) shows their symmetry behavior under the chirality exchange of the decaying and final charged lepton, while the subscript ($jk$) defines the specific mixing matrix elements that multiply the coupling constant, which in some cases are permuted due to renaming of dummy indices to simplify the final expression.\\
We have already distinguished the new contributions that a possible Majorana nature of neutrinos would have on these parameters with the blue square brackets. Finally, due to the neutrino mass dependence, we have new phase-space structures and the possibility to discriminate between a Dirac or Majorana nature of neutrinos, as well as the existence of possible new physics. In the next subsections we shall compare the final Dirac and Majorana distributions, discussing their common features as well as their main differences.

\subsection{Common Features}
\label{subsec:4.1}
\begin{figure}[tbp]
\centering
\begin{subfigure}[]{0.49\textwidth}
   \includegraphics[width=1\linewidth]{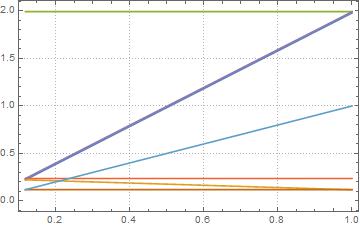}
   \caption{$\tau\rightarrow\mu\nu\overline{\nu} \quad (x_0=0.118508)$}
   \label{fig:Ng1} 
\end{subfigure}
\begin{subfigure}[]{0.49\textwidth}
   \includegraphics[width=1\linewidth]{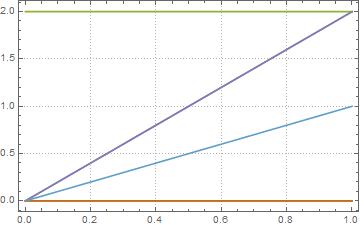}
   \caption{$\tau\rightarrow e\nu\overline{\nu}\quad (x_0=0.000574)$}
   \label{fig:Ng2}
\end{subfigure}
\begin{subfigure}[]{0.75\textwidth}
   \includegraphics[width=1\linewidth]{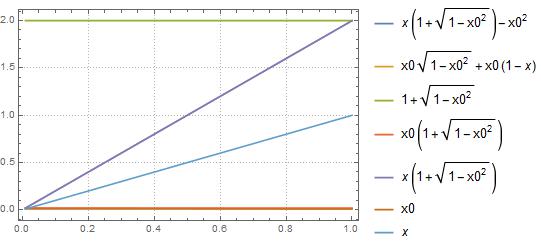}
   \caption{$\mu\rightarrow e\nu\overline{\nu}\quad (x_0=0.009653)$}
   \label{fig:Ng3}
\end{subfigure}
\caption{Different phase space structures for the leptonic differential decay rate ($x_0<x<1$).}
\label{fig:phasespace}
\end{figure}

 From the final distribution result it is straightforward that both, Dirac and Majorana distributions, have the same $x_0$ and $x$ dependence, as well as the linear and quadratic neutrino mass suppression. Thus the following phase-space analysis and neutrino mass suppression effects are valid for either of them.\\
 We shall focus on the suppression due only to the masses and mixings, considering just one extra heavy neutrino for simplicity. Of course there will be other suppression factors such as the specific phase-space structure dependence and the explicit form of the new parameters, but this will not be the main contribution, so we are not taking them into account in this analysis, we will study them later.\\
$\bullet$ If the heavy neutrino is forbidden by kinematics, then only the light neutrinos will be produced as final states and thus the suppression will be really high. Considering the light neutrino masses to be of order $\mathcal{O}$(eV) and the decaying particle mass of order $\mathcal{O}$($10^9$eV), then the new contributions with neutrino mass dependence will be suppressed by a factor of $\sim 10^{-9}$ for the linear mass terms and $\sim 10^{-18}$ for the quadratic one. Both of them out of the scope of near-future experiments.\\
$\bullet$ In addition, the absence of heavy neutrinos would have an impact on the non-unitarity of the mixing matrix, leading to small deviations from unity once the squared mixing matrix elements are summed over all the energetically allowed neutrinos.\\
$\bullet$ In contrast, if the heavy neutrinos are kinematically accessible, then the suppression of the terms with explicit neutrino mass dependence will change, depending on the heavy neutrino mass and its mixing with the active and sterile sector.\\
Following our notation, the mixing matrix elements $ U_{\ell j}$ and $V_{\ell j}$ are suppressed depending on the neutrino mass, as shown in table \ref{tab:suppression1}.\\
\begin{table}[tbp]
\centering
\begin{tabular}{|c|c|c|}
\hline
  Neutrino & $U_{\ell j}$& $V_{\ell j}$\\ [5pt]
\hline
Light ($j\leq3$)& Not suppressed & Suppressed\\ [10pt]
\hline
Heavy ($j\geq4$)& Suppressed &  Not suppressed \\ [10pt]
\hline
\end{tabular}
\caption{\label{tab:suppression1} Suppression of the mixing matrix elements.}
\end{table}
Considering the experimental constraints on an invisible heavy neutrino $\nu_4$ (table \ref{tab:suppression2}), obtained from different experimental sources and some reasonable phenomenological assumptions ($\nu_4$ decays and its lifetime), see e.g. \cite{45,46}~\footnote{In this table we focus on the mass intervals that can most effectively be probed in the Michel decays. See, e. g., Ref. \cite{Kim:2018uht} for bounds on sub-eV scale sterile neutrinos and its connection to anomalies in short-baseline neutrino oscillation experiments.}, we can estimate the suppression of the neutrino mass dependent terms compared with the ones without this dependence (standard Michel distribution).\\
\begin{table}[tbp]
\centering
\begin{tabular}{|c|c|c|}
\hline
  Neutrino &Mass (MeV) &Mixing $|U_{\ell4}|^2$\\ [5pt]
\hline
Heavy ($\ell= e$)&0.001 - 0.45&$10^{-3}$  \\ [10pt]

&10 - 55&$10^{-8}$  \\ [10pt]
&135 - 350&$10^{-6}$  \\ [10pt]
\hline
Heavy ($\ell= \mu$)&10 - 30&$10^{-4}$  \\ [10pt]
&70 - 300&$10^{-5}$  \\ [10pt]
&175 - 300&$10^{-8}$  \\ [10pt]

\hline
Heavy ($\ell= \tau$)&100 - $1.2\times10^{3}$&$10^{-7} - 10^{-3}$  \\ [10pt]

&$1\times10^{3}-60\times10^{3}$&$10^{-5} - 10^{-3}$  \\ [10pt]
\hline
\end{tabular}
\caption{ Experimental constraints on a heavy neutrino mixing \cite{45,46}. \label{tab:suppression2}}
\end{table}
Thus it is possible to have one or two heavy neutrinos in the final states and each term will be suppressed by some of the mixing matrix elements, depending on the specific form of the $[f^n_{lm}]_{jk}$ coupling constants.\\
For this rough estimation we will consider both cases (one and two final heavy neutrinos) and the mixing suppression as $|U_{\ell4}|^2$ for one final heavy neutrino and $|U_{\ell4}|^2|U_{\ell^{'}4}|^2$ for two heavy final-state neutrinos. With these considerations we can summarize the results of the suppression of the neutrino mass dependent terms in table \ref{tab:suppression3}.\\
\begin{table}[tbp]
\centering
\begin{tabular}{|c|c|c|c|c|}
\hline
  Neutrino &Mass (MeV) &\makecell{Mixing \\ Suppression}& \makecell{Linear Term \\ Suppression ($m_\nu$)}& \makecell{Quadratic Term \\ Suppression ($m_\nu^2$)}\\ [5pt]
\hline
Light (2)& $1\times10^{-6}$ & --- & $10^{-9}$ & $10^{-18}$ \\ [10pt]
\hline
\makecell{Heavy (1) \\ ($\ell=e$)} &0.001 - 0.45 & $10^{-3}$ &$10^{-9} - 10^{-7}$ & $10^{-18} - 10^{-16}$ \\ [10pt]

&10 - 55&$10^{-8}$ &$10^{-10}$ & $10^{-19}$\\ [10pt]
&135 - 350&$10^{-6}$ &$10^{-7}$ & $10^{-16}$\\ [10pt]
\hline
\makecell{Heavy (1) \\ ($\ell=\mu$)}&10 - 30&$10^{-4}$ & $10^{-6}$& $10^{-15}$ \\ [10pt]

&70 - 300&$10^{-5}$ &$10^{-7} - 10^{-6}$ & $10^{-16} - 10^{-15}$ \\ [10pt]
&175 - 300&$10^{-8}$ & $10^{-9}$& $10^{-18}$\\ [10pt]
\hline

\makecell{Heavy (1) \\ ($\ell=\tau$)}&$100 - 1.2\times10^{3}$&$10^{-7} - 10^{-3}$  & $10^{-8} - 10^{-3}$& $10^{-18} - 10^{-12}$ \\ [10pt]

&$1\times10^{3}-60\times10^{3}$&$10^{-5} - 10^{-3}$ & $10^{-5} - 10^{-3}$ & $10^{-14} - 10^{-12}$ \\ [10pt]
\hline
\makecell{Heavy (2) \\ ($\mu\rightarrow e N N$)} &10 - 30 &$10^{-12}$ & $10^{-14}$& $10^{-16}$ \\ [10pt]

&175 - 300&$10^{-14} - 10^{-11}$ &$10^{-15} - 10^{-12}$ & $10^{-16} - 10^{-13}$\\ [10pt]
\hline
\makecell{Heavy (2) \\ ($\tau\rightarrow e N N$)}&135 - 350&$10^{-13} - 10^{-9}$ & $10^{-14} - 10^{-10}$ & $10^{-14} - 10^{-10}$ \\ [10pt]

\hline

\makecell{Heavy (2) \\ ($\tau\rightarrow \mu N N$)} &100 - 300&1$0^{-12} - 10^{-8}$ & $10^{-13} - 10^{-9}$ & $10^{-14} - 10^{-10}$ \\ [10pt]
&175 - 350&$10^{-15} - 10^{-11}$ & $10^{-16} - 10^{-12}$ & $10^{-16} - 10^{-12}$ \\ [10pt]

\hline
\end{tabular}
\caption{Estimation of neutrino mass dependent terms suppression. \label{tab:suppression3}}
\end{table}
The mean life of the muon and tau has been measured to a precision of order $10^{-6}$ and $10^{-3}$ respectively \cite{1}. In order to make new precision tests, the most recent and future experiments are working hard to increase this sensitivity. From table \ref{tab:suppression3}, sadly, almost all of the new contributions are really suppressed and out of reach at the near future experiments, but a few of them are just around the current precision limit and could be measurable in current and forthcoming experiments.\\
$\bullet$ Specifically, in the case of one final heavy neutrino with a mass around $10^2-10^3$ MeV\footnote{We note that, in this mass range, the limit on $|U_{\ell 4}|^2$ has recently been slightly improved in ref.~\cite{BaBar:2022cqj}.}, the linear term suppression could be low enough to make sizeable distortions in the differential decay rate, specially in tau decays, where the Majorana or Dirac neutrino nature can be tested, as well as the underlying Lorentz structure and possible new physics.\\
$\bullet$ Finally since the maximum measurable effect of a neutrino mass in the decay rate would be of order $10^{-3}$ for $\tau$-decay and $10^{-6}$ for $\mu$-decay, as shown in table \ref{tab:suppression3}, it is important for the precision electroweak tests to consider all radiative corrections and their sub-leading effects.\\
These effects take into account radiative QED corrections,
higher-order electroweak corrections and the non-local structure of the $W$ propagator, all of these within the SM framework, where the corrections to the total decay rate are well-known at this precision level \cite{24,25,26}.\\
These corrections can also be analyzed at the level of differential decay rate, specifically the most recent corrections induced by the $W$-boson propagator to the differential rates of the leptonic decay of a polarized muon and tau lepton and the numerical effect of these corrections are discussed in \cite{47} and analyzed in appendix \ref{appendix:A} for the specific Michel distribution, where we also introduce the result for this correction taking into account a final charged-lepton polarization (for a recent related proposal see \cite{Bodrov:2022mbd}).\\
It is important to emphasize that we can safely employ these radiative corrections, calculated in the SM limit $|f_{LL}^V|=1$, in order to measure with high precision the Michel parameters, since, to a high degree of accuracy, the current experimental information is consistent with a $V-A$ structure, so possible deviations are expected to be very small and can therefore be treated at the tree level, making the SM radiative corrections the main higher-order contributions. The interested reader is addressed to ref.~\cite{48} for a helpful discussion.\\We also summarize the main numerical contributions, including hadronic corrections, in table \ref{tab:radcor},
\begin{table}[tbp]
\centering
\begin{tabular}{|c|c|c|}
\hline
  Radiative Corrections and & Numerical Effect ($\mu$-decay) & Numerical Effect ($\tau$-decay)\\
  Finite Mass Effects& & \\ [5pt]
\hline
Electroweak & $(3/5)(m_\mu^2/M_W^2)\sim 1.0\times10^{-6}$ & $(3/5)(m_\tau^2/M_W^2)\sim 2.9\times10^{-4}$\\ [10pt]
\hline
QED &  $\mathcal{O}(\alpha)\sim 10^{-3}$& $\mathcal{O}(\alpha)\sim 10^{-3}$ \\ [10pt]
\hline
Hadronic & $\mathcal{O}(\alpha^2/\pi^2)\sim 10^{-5}$ & $\mathcal{O}(\alpha^2/\pi^2)\sim 10^{-5}$\\ [10pt]
\hline

\end{tabular}
\caption{ Main numerical effects of radiative corrections. \label{tab:radcor}}
\end{table}
where the subleading contributions of these corrections are of order $\mathcal{O}(10^{-11}-10^{-7})$, which are out of experimental reach in the foreseeable future. \\
From table \ref{tab:radcor} it is evident that the three main radiative corrections must be taken into account in the muon decay analysis, since the smallest one of them is of the same magnitude as the present experimental relative uncertainty of the muon decay rate.\\
$\bullet$ In principle for tau decays, due to the current precision achieved, the QED and electroweak corrections are the most important. The hadronic corrections are not needed, since it would imply a correction up to 1-10\% to something that has not yet been measured. \\
Another suppression effect is due to the explicit structure of the differential decay rate's phase space. Both Dirac and Majorana distributions have the same $x_0$ and $x$ dependence, thus the following analysis is valid for either of them.\\
We shall focus on the phase space terms proportional to and quadratic on neutrino masses. From the kinematics analysis of 1$\rightarrow$ 3 processes; we know that $x_0<x<1$. With these limits, we can study the behavior of the new phase space structures, see figure \ref{fig:phasespace}.\\
$\bullet$ As we can see, the phase space effects are greater at a large value of $x$ in almost all the cases and neither of them can be bigger than two units. Thus the possible Majorana or Dirac effects are more suppressed at lower $x$ and are slightly increased for a larger $x$.\\
Obviously if we consider the specific form of the new parameters (Dirac or Majorana) there will be possible effects of constructive or destructive interference, depending on the value of the $f_{lm}^n$ coupling constants, but in this analysis we do not discuss any of these effects, we just focus on neutrino masses and phase space contributions.\\
Finally it is important to emphasize again that we are neglecting terms with neutrino mass dependence in the phase space integration, so actually the upper $x$ limit will be less than one unit, even so, for greater $x$, the neutrinos mass effects are slightly increased.\\
$\bullet$ With these results, the presence of a heavy Dirac or Majorana neutrino will make measurable distortions in the differential decay rate, specially in the case of a $\tau$-decay with one heavy final-state neutrino with a mass around $10^2 - 10^3$ MeV and in the $x\approx 1$ phase space zone, i.e., where the final charged lepton reaches its maximum energy.

\subsection{Main Differences}
\label{subsec:4.2}
As discussed before, the only difference between the Dirac and Majorana differential decay rate is the interference $Majorana$ $term$. This new contribution leads to some changes in every $Dirac$ parameter, where depending on the new physics involved, it could generate a measurable distortion. Thus the study of this interference term could be a potential way to define the neutrino nature as well as the presence of new physics.\\
In our analysis we define this new contribution with the help of the flag parameter $\epsilon$, being $\epsilon=0$ for the Dirac case and $\epsilon=1$ for the Majorana one. Since the difference between these two cases is precisely this $Majorana$ term, we show the full and explicit dependence on the coupling constant of each parameter due to this Majorana contribution in table \ref{tab:dirmajo1}. \\
$\bullet$ We separate explicitly the parameters that characterize the differential decay rate with no,  linear and  quadratic neutrino mass dependence, respectively. This separation allows us to quickly identify the suppression that each parameter would have due to the neutrino masses and discuss their properties in some helpful examples.\\ 
$\bullet$ It is important to emphasize that this explicit differentiation of the Majorana contribution and the neutrino mass dependence terms, makes much easier the implementation of our results in specific model-dependent theories, where the effects of neutrino mass and its specific nature could be seen immediately.  
\begin{table}[]
\centering
\resizebox{\textwidth}{!}{
\begin{tabular}{|c|c|}
\hline
  \textbf{Term}& \textbf{Coupling Dependence}  \\ [5pt]
\hline
\multicolumn{2}{|c|}{\textbf{No Neutrino Mass Dependence}}\\ [5pt]
\hline
 $(I)^M_{jk}$& \begin{tabular}{@{}c@{}} $\frac{1}{8}\Big[\color{black}12(f_{LR}^{T})_{jk}(f_{LR}^{S})^*_{kj}+12(f_{LR}^{T})_{jk}(f_{LR}^{T})^*_{kj}+8(f_{RL}^{V})_{jk}(f_{RL}^{V})^*_{kj}-(f_{LR}^{S})_{jk}(f_{LR}^{S})^*_{kj}  $\\
 $+8(f_{LL}^{S})_{jk}(f_{LL}^{V})^*_{kj}+(L\leftrightarrow R)\Big]\color{black}$\end{tabular}\\ 
 \hline
 $(\rho)^M_{jk}$& \begin{tabular}{@{}c@{}}  $\frac{3}{16}\Big[\color{black} -(f_{LR}^{S})_{jk}(f_{LR}^{S})^*_{kj}+4(f_{LR}^{S})_{jk}(f_{LR}^{T})^*_{kj}+4(f_{LL}^{S})_{jk}(f_{LL}^{V})^*_{kj}-4(f_{LR}^{T})_{jk}(f_{LR}^{T})^*_{kj}+(L\leftrightarrow R) \Big]\color{black}$ \end{tabular} \\ 
 \hline
$(\xi)^M_{jk} $&\begin{tabular}{@{}c@{}}$\color{black}-(f_{RR}^{S})_{jk}(f_{RR}^{V})^*_{kj}+\frac{17}{2}(f_{LR}^{T})_{jk}(f_{LR}^{T})^*_{kj}+\frac{1}{2}(f_{LR}^{S})_{jk}(f_{LR}^{T})^*_{kj}+3(f_{LR}^{V})_{jk}(f_{LR}^{V})^*_{kj}$\\ $+\frac{5}{8}(f_{LR}^{S})_{jk}(f_{LR}^{S})^*_{kj}-(L\leftrightarrow R) $ \end{tabular}   \\ 
 \hline
$(\xi\delta)^M_{jk}$ &  \begin{tabular}{@{}c@{}} $\frac{3}{4}\Big[\color{black} -(f_{RR}^{S})_{jk}(f_{RR}^{V})^*_{kj}+(f_{LR}^{T})_{jk}(f_{LR}^{T})^*_{kj}-(f_{LR}^{S})_{jk}(f_{LR}^{T})^*_{kj}+\frac{1}{4}(f_{LR}^{S})_{jk}(f_{LR}^{S})^*_{kj}-(L\leftrightarrow R)\Big]\color{black}$ \end{tabular} \\ 
 \hline
$(\eta)^M_{jk}$ &  \begin{tabular}{@{}c@{}}$\frac{1}{8}\Big[\color{black}4(f_{LR}^{S})_{jk}(f_{RL}^{V})^*_{kj}+24(f_{LR}^{T})_{jk}(f_{RL}^{V})^*_{kj}+(f_{LL}^{S})_{jk}(f_{RR}^{S})^*_{kj}+4(f_{LL}^{V})_{jk}(f_{RR}^{V})^*_{kj}+(L\leftrightarrow R)\Big]\color{black}$ \end{tabular} \\ 
 \hline
$(\xi^{'})^M_{jk}$ &  \begin{tabular}{@{}c@{}}$\color{black}(f_{LL}^{S})_{jk}(f_{LL}^{V})^*_{kj}+\frac{3}{2}(f_{LR}^{S})_{jk}(f_{LR}^{T})^*_{kj}+\frac{3}{2}(f_{LR}^{T})_{jk}(f_{LR}^{T})^*_{kj}+(f_{LR}^{V})_{jk}(f_{LR}^{V})^*_{kj}$\\$-\frac{1}{8}(f_{LR}^{S})_{jk}(f_{LR}^{S})^*_{kj}-(L\leftrightarrow R)$ \end{tabular} \\ 
 \hline
 $(\xi^{''})^M_{jk}$&  \begin{tabular}{@{}c@{}}$\color{black}\frac{1}{2}(f_{LR}^{S})_{jk}(f_{LR}^{T})^*_{kj}+\frac{17}{2}(f_{LR}^{T})_{jk}(f_{LR}^{T})^*_{kj}+(f_{LL}^{S})_{jk}(f_{LL}^{V})^*_{kj}+\frac{5}{8}(f_{RL}^{S})_{jk}(f_{RL}^{S})^*_{kj}$\\$+3(f_{RL}^{V})_{jk}(f_{RL}^{V})^*_{kj}+(L\leftrightarrow R) $ \end{tabular} \\ 
 \hline
$(\eta^{''})^M_{jk}$ &  \begin{tabular}{@{}c@{}}$\frac{1}{2}\Big[\color{black} 3(f_{LR}^{S})_{jk}(f_{RL}^{V})^*_{kj}+18(f_{LR}^{T})_{jk}(f_{RL}^{V})^*_{kj}-\frac{1}{4}(f_{LL}^{S})_{jk}(f_{RR}^{S})^*_{kj}-(f_{LL}^{V})_{jk}(f_{RR}^{V})^*_{kj}+(L\leftrightarrow R)\Big]\color{black}$ \end{tabular} \\ 
 \hline
 $(\frac{\alpha^{'}}{\mathcal{A}})^M_{jk}$&  \begin{tabular}{@{}c@{}} $\frac{1}{2}(f_{RL}^{V})^*_{jk}((f_{LR}^{S})_{kj}+6(f_{LR}^{T})_{kj})-(L\leftrightarrow R)$ \end{tabular} \\ 
 \hline
$(\frac{\beta^{'}}{\mathcal{A}})^M_{jk}$ &  \begin{tabular}{@{}c@{}}$-\frac{1}{2} (f_{LL}^{V})_{jk}(f_{RR}^{V})^*_{kj}-\frac{1}{8}(f_{LL}^{S})_{jk}(f_{RR}^{S})^*_{kj}$ \end{tabular} \\ 
\hline
\multicolumn{2}{|c|}{\textbf{Linear Neutrino Mass Dependence}}\\ [5pt]
\hline
 $(\kappa^{\pm}_L)^M_{jk}$&  \begin{tabular}{@{}c@{}} $\color{black}-2(f_{LL}^{V})_{kj}(f_{LR}^{V})^*_{jk}-\frac{1}{2}(f_{LL}^{S})_{kj}(f_{LR}^{S})^*_{jk}+3(f_{LL}^{S})_{kj}(f_{LR}^{T})^*_{jk}+2(f_{LL}^{V})_{jk}(f_{LR}^{S})^*_{kj}$\\$-(f_{LL}^{S})_{jk}(f_{LR}^{V})^*_{kj}\pm(L\leftrightarrow R)$\end{tabular} \\
\hline
$(\kappa^{\pm}_R)^M_{jk}$&  \begin{tabular}{@{}c@{}} $\color{black}-2(f_{RR}^{V})_{kj}(f_{LR}^{V})^*_{jk}-\frac{1}{2}(f_{RR}^{S})_{kj}(f_{LR}^{S})^*_{jk}+3(f_{RR}^{S})_{kj}(f_{LR}^{T})^*_{jk}+2(f_{RR}^{V})_{jk}(f_{LR}^{S})^*_{kj}$\\$-(f_{RR}^{S})_{jk}(f_{LR}^{V})^*_{kj}\pm(L\leftrightarrow R)$\end{tabular} \\
\hline
$(\lambda^{\pm}_L)^M_{jk}$ &  \begin{tabular}{@{}c@{}}$ -2(f_{LL}^{V})_{kj}(f_{LR}^{V})^*_{jk}+\frac{1}{2}(f_{LL}^{S})_{kj}(f_{LR}^{S})^*_{jk}+(f_{LL}^{S})_{kj}(f_{LR}^{T})^*_{jk}+4(f_{LL}^{V})_{jk}(f_{LR}^{T})^*_{kj}$\\$-(f_{LL}^{S})_{jk}(f_{LR}^{V})^*_{kj}\pm(L \leftrightarrow R)$\end{tabular} \\
\hline
$(\lambda^{\pm}_R)^M_{jk}$ &  \begin{tabular}{@{}c@{}}$ -2(f_{RR}^{V})_{kj}(f_{LR}^{V})^*_{jk}+\frac{1}{2}(f_{RR}^{S})_{kj}(f_{LR}^{S})^*_{jk}+(f_{RR}^{S})_{kj}(f_{LR}^{T})^*_{jk}+4(f_{RR}^{V})_{jk}(f_{LR}^{T})^*_{kj}$\\$-(f_{RR}^{S})_{jk}(f_{LR}^{V})^*_{kj}\pm(L \leftrightarrow R)$\end{tabular} \\
\hline
\multicolumn{2}{|c|}{\textbf{Quadratic Neutrino Mass Dependence}}\\ [5pt]
\hline
$(C^{\pm})^M_{jk}$ &  \begin{tabular}{@{}c@{}} $(f_{LL}^{V})_{jk}(f_{LL}^{V})^*_{kj}+\frac{1}{4}(f_{LL}^{S})_{jk}(f_{LL}^{S})^*_{kj}+(f_{RL}^{S})_{jk}(f_{RL}^{V})^*_{kj}+6(f_{RL}^{V})_{jk}(f_{RL}^{T})^*_{kj}\pm(L\leftrightarrow R)$ \end{tabular} \\
\hline
 $(C^{'\pm})^M_{jk}$&  \begin{tabular}{@{}c@{}} $\frac{1}{4}(f_{LL}^{S})_{jk}(f_{LL}^{S})^*_{kj}+(f_{LL}^{V})_{jk}(f_{LL}^{V})^*_{kj}-(f_{RL}^{S})_{jk}(f_{RL}^{V})^*_{kj}-6(f_{RL}^{V})_{jk}(f_{RL}^{T})^*_{kj}\pm(L \leftrightarrow R)$\end{tabular} \\
\hline
$(J^{+})^M_{jk}$ &  \begin{tabular}{@{}c@{}} $\frac{1}{4}(f_{LR}^{S})_{kj}(f_{RL}^{S})^*_{jk}+\frac{1}{2}(f_{LR}^{S})_{kj}(f_{RL}^{T})^*_{jk}+\frac{1}{2}(f_{LR}^{T})_{kj}(f_{RL}^{S})^*_{jk}+5(f_{LR}^{T})_{kj}(f_{RL}^{T})^*_{jk}$\\$+2(f_{LR}^{V})_{jk}(f_{RL}^{V})^*_{kj}$\end{tabular} \\
\hline
 $(H^{+})^M_{jk}$&  \begin{tabular}{@{}c@{}} $2(f_{LL}^{V})_{jk}(f_{RR}^{S})^*_{kj}-\frac{1}{4}(f_{LR}^{S})_{jk}(f_{RL}^{S})^*_{kj}+3(f_{LR}^{T})_{jk}(f_{RL}^{S})^*_{kj}+3(f_{LR}^{T})_{jk}(f_{RL}^{T})^*_{kj}$\\$+2(f_{LR}^{V})_{jk}(f_{RL}^{V})^*_{kj}+(L\leftrightarrow R)$\end{tabular} \\
\hline
\end{tabular}}
\caption{Majorana term for the general coupling case. \label{tab:dirmajo1}}
\end{table}
From table \ref{tab:dirmajo1}, it is straightforward that the discrepancy between the Dirac and Majorana cases depends in a non-trivial way on the coupling constants, specifically the new physics ones. This dependence could help us to characterize the underlying Lorentz structure of a theory, as we shall see in the next examples.\\
First of all, if no new-physics is present, i.e., in the SM case ($|f_{LL}^{V}|=1$), we obtained the results displayed in table \ref{tab:dirmajo2} for the Majorana term.
\begin{table}[]
\centering
\begin{tabular}{|c|c|c|c|}
\hline
  \textbf{Term}& \textbf{Coupling Dependence}&\textbf{Term}& \textbf{Coupling Dependence}  \\ [5pt]
\hline
\multicolumn{2}{|c|}{\textbf{No Neutrino Mass Dependence}}&\multicolumn{2}{|c|}{\textbf{Linear Neutrino Mass Dependence}}\\ [5pt]
\hline
 $(I)^M_{jk}$& \begin{tabular}{@{}c@{}} 0 \end{tabular}&$(\kappa^{\pm}_L)^M_{jk}$&  \begin{tabular}{@{}c@{}} 0 \end{tabular}\\ 
 \hline
 $(\rho)^M_{jk}$& \begin{tabular}{@{}c@{}}  0 \end{tabular}& $(\kappa^{\pm}_R)^M_{jk}$&  \begin{tabular}{@{}c@{}} 0 \end{tabular} \\ 
 \hline
$(\xi)^M_{jk}$ &\begin{tabular}{@{}c@{}} 0 \end{tabular}   &$(\lambda^{\pm}_L)^M_{jk}$ &  \begin{tabular}{@{}c@{}} 0 \end{tabular}\\ 
 \hline
$(\xi\delta)^M_{jk}$ &  \begin{tabular}{@{}c@{}} 0 \end{tabular}&$(\lambda^{\pm}_R)^M_{jk}$ &  \begin{tabular}{@{}c@{}} 0 \end{tabular} \\ 
 \hline
$(\eta)^M_{jk}$ &  \begin{tabular}{@{}c@{}} 0 \end{tabular} & \multicolumn{2}{|c|}{\textbf{Quadratic Neutrino Mass Dependence}}\\ 
 \hline
$(\xi^{'})^M_{jk}$ &  \begin{tabular}{@{}c@{}} 0 \end{tabular}& $(C^{\pm})^M_{jk}$ &  \begin{tabular}{@{}c@{}} $(f_{LL}^{V})_{jk}(f_{LL}^{V})^*_{kj}$ \end{tabular} \\ 
 \hline
 $(\xi^{''})^M_{jk}$&  \begin{tabular}{@{}c@{}} 0 \end{tabular}& $(C^{'\pm})^M_{jk}$&  \begin{tabular}{@{}c@{}} $(f_{LL}^{V})_{jk}(f_{LL}^{V})^*_{kj}$\end{tabular}  \\ 
 \hline
$(\eta^{''})^M_{jk}$ &  \begin{tabular}{@{}c@{}} 0 \end{tabular}&$(J^{+})^M_{jk}$ &  \begin{tabular}{@{}c@{}} 0\end{tabular} \\ 
 \hline
 $(\frac{\alpha^{'}}{\mathcal{A}})^M_{jk}$&  \begin{tabular}{@{}c@{}} 0 \end{tabular}& $(H^{+})^M_{jk}$&  \begin{tabular}{@{}c@{}} 0 \end{tabular} \\ 
 \hline
$(\frac{\beta^{'}}{\mathcal{A}})^M_{jk}$ &  \begin{tabular}{@{}c@{}} 0 \end{tabular} & &\\ 
\hline
\end{tabular}
\caption{ SM Case $|f_{LL}^{V}|=1$. \label{tab:dirmajo2}}
\end{table}
This result shows explicitly the $Kaiser$ $Theorem$\footnote{Kaiser theorem can be avoided if neutrinos momenta are not integrated out, unfortunately since neutrinos are not detected we can not measure their momenta directly. Alternatives of indirect measurements are discussed in \cite{50} but not applicable in this work.} \cite{49}, that states: "For massless neutrinos in a world where all weak currents are left-handed, there is no distinction between a two-component Dirac (i.e., Weyl)
neutrino and a Majorana neutrino. Once the mass is non-zero, however, then no matter how small it is, a Dirac and a Majorana neutrino
are different." \\
Indeed from table \ref{tab:dirmajo2}, we see that in the SM case, there is no difference between a Dirac and a Majorana neutrino for the terms with no neutrino mass dependence (also for the linear neutrino mass dependence), but, even in the SM case, if neutrino masses are taken  into account, we find a substantial difference between the Dirac and Majorana cases, specifically in the quadratic neutrino mass dependence.
Unfortunately these terms are really suppressed, 
as shown in table \ref{tab:suppression3}.\\
Finally we can check the consistency of our calculation by reproducing well-known results. It is possible to show that, in the limit $m_\nu\rightarrow 0$, our final result (eq.(\ref{completepararesults})) perfectly agrees with the ones obtained by Paul Langacker and David London \cite{42}, where our coupling constants are related by:
\begin{equation}
    \begin{split}
        &\bullet g^{LL}_{ij}=(f^{V}_{LL})_{jk}\quad\quad\bullet g^{++}_{ij}=(f^{S}_{LR})_{jk}\quad\quad\bullet g^{T+}_{ij}=(f^{T}_{LR})_{jk}\\
        &\bullet g^{LR}_{ij}=(f^{V}_{LR})_{jk}\quad\quad\bullet g^{+-}_{ij}=(f^{S}_{LL})_{jk}\quad\quad\bullet g^{T-}_{ij}=(f^{T}_{RL})_{jk}\\
         &\bullet g^{RL}_{ij}=(f^{V}_{RL})_{jk}\quad\quad\bullet g^{-+}_{ij}=(f^{S}_{RR})_{jk}\\
          &\bullet g^{RR}_{ij}=(f^{V}_{RR})_{jk}\quad\quad\bullet g^{--}_{ij}=(f^{S}_{RL})_{jk}
    \end{split}
\end{equation}
The comparison with the general result of Shrock \cite{51} is a bit more subtle, because our coupling constants are related by a more general Fierz transformation, since he used the $charged$ $retention$ form of the Hamiltonian.\\
Using the relations between constants, derived in Appendix \ref{appendix:B}, we can compare our results including neutrino masses effects and a possible Majorana nature of neutrinos.\footnote{Remember that we are just dealing with neutrino masses up to the matrix element calculation, neglecting them in the phase space treatment as a first and good approximation.} Our results agree with the ones obtained by Shrock, except for the linear neutrino mass terms that multiplies the $m_a$ factor of his equations (21) and (22), and the $m_b$ factor of his equation (26), which seem to be a typo in his work, and the quadratic neutrino mass term of equation (26), which is a very suppressed contribution anyway.\\
It is important to note that their calculations for the Majorana neutrino case were done using the coupling constants correspondence at the Hamiltonian level, through Fierz transformation, as shown explicitly in \cite{42}. This method makes the calculation faster, but does not show clearly the fundamental difference that the Majorana contribution could have in model-dependent theories. Furthermore, the parametrization used in \cite{51} is not commonly employed in experimental analysis, where the Michel (PDG) parametrization is currently the most useful. \\
On the other hand we distinguish the Majorana contribution by explicit calculation of the $Majorana$ $term$. Also, we rename dummy indices wherever possible in order to make our expressions clearer and easier to use for general applications and write all our results in the PDG parametrization form, which is implemented by many experiments.\\
Finally we show the fast implementation of our results in a well-studied model-dependent theory which includes finite neutrino mass effects. This theory is the one obtained within the framework of the effective weak interaction Hamiltonian with the ($V\pm A$) currents \cite{12,52}. For the presence of $V\pm A$ currents, keeping only the leading order terms, the Majorana terms are shown in table \ref{tab:dirmajo7}.\\
\begin{table}[]
\centering
\begin{tabular}{|c|c|c|c|}
\hline
  \textbf{Term}& \textbf{Coupling Dependence}&\textbf{Term}& \textbf{Coupling Dependence}  \\ [5pt]
\hline
\multicolumn{2}{|c|}{\textbf{No Neutrino Mass Dependence}}&\multicolumn{2}{|c|}{\textbf{Linear Neutrino Mass Dependence}}\\ [5pt]
\hline
 $(I)^M_{jk}$& \begin{tabular}{@{}c@{}} 0 \end{tabular}&$(\kappa^{\pm}_L)^M_{jk}$&  \begin{tabular}{@{}c@{}} $-2(f_{LL}^{V})_{kj}(f_{LR}^{V})^*_{jk}$ \end{tabular}\\ 
 \hline
 $(\rho)^M_{jk}$& \begin{tabular}{@{}c@{}}  0 \end{tabular}& $(\kappa^{\pm}_R)^M_{jk}$&  \begin{tabular}{@{}c@{}} $\mp 2(f_{LL}^{V})_{kj}(f_{RL}^{V})^*_{jk}$ \end{tabular} \\ 
 \hline
$(\xi)^M_{jk}$ &\begin{tabular}{@{}c@{}} 0 \end{tabular}   &$(\lambda^{\pm}_L)^M_{jk}$ &  \begin{tabular}{@{}c@{}} $-2(f_{LL}^{V})_{kj}(f_{LR}^{V})^*_{jk}$ \end{tabular}\\ 
 \hline
$(\xi\delta)^M_{jk}$ &  \begin{tabular}{@{}c@{}} 0 \end{tabular}&$(\lambda^{\pm}_R)^M_{jk}$ &  \begin{tabular}{@{}c@{}} $\mp 2(f_{LL}^{V})_{kj}(f_{RL}^{V})^*_{jk}$ \end{tabular} \\ 
 \hline
$(\eta)^M_{jk}$ &  \begin{tabular}{@{}c@{}} $(f_{LL}^{V})_{jk}(f_{RR}^{V})^*_{kj}$ \end{tabular} & \multicolumn{2}{|c|}{\textbf{Quadratic Neutrino Mass Dependence}}\\ 
 \hline
$(\xi^{'})^M_{jk}$ &  \begin{tabular}{@{}c@{}} 0 \end{tabular}& $(C^{\pm})^M_{jk}$ &  \begin{tabular}{@{}c@{}} $(f_{LL}^{V})_{jk}(f_{LL}^{V})^*_{kj}$ \end{tabular} \\ 
 \hline
 $(\xi^{''})^M_{jk}$&  \begin{tabular}{@{}c@{}} 0 \end{tabular}&$(C^{'\pm})^M_{jk}$&  \begin{tabular}{@{}c@{}} $(f_{LL}^{V})_{jk}(f_{LL}^{V})^*_{kj}$\end{tabular}  \\ 
 \hline
$(\eta^{''})^M_{jk}$ &  \begin{tabular}{@{}c@{}} $-(f_{LL}^{V})_{jk}(f_{RR}^{V})^*_{kj}$ \end{tabular}&$(J^{+})^M_{jk}$ &  \begin{tabular}{@{}c@{}} 0\end{tabular} \\ 
 \hline
 $(\frac{\alpha^{'}}{\mathcal{A}})^M_{jk}$&  \begin{tabular}{@{}c@{}} 0 \end{tabular}& $(H^{+})^M_{jk}$&  \begin{tabular}{@{}c@{}} 0 \end{tabular} \\ 
 \hline
$(\frac{\beta^{'}}{\mathcal{A}})^M_{jk}$ &  \begin{tabular}{@{}c@{}} $-\frac{1}{2}(f_{LL}^{V})_{jk}(f_{RR}^{V})^*_{kj}$ \end{tabular} & &\\ 
\hline
\end{tabular}
\caption{ Majorana term for $V\pm A$ currents. \label{tab:dirmajo7}}
\end{table}
Then, if we focus now only on the component of the transverse polarization perpendicular to the decay plane ($F_{T_2}$ and $F^{'}_{T_2}$) up to linear neutrino mass effects, taking into account the Dirac ($\epsilon=0$) and Majorana ($\epsilon=1$) cases and considering $x_0<<1$ we can approximate the result to:
\begin{equation}
\begin{split}
    F_{T_2}(x)+F_{T_2}'(x)=&\frac{x}{3m_\ell}\bigg\{\epsilon\Big[m_\ell\operatorname{Im}(f_{LL}^{V})^*_{jk}(f_{RR}^{V})_{kj}-3m_j\operatorname{Im}(f_{LL}^{V})^*_{jk}(f_{RL}^{V})_{kj}\Big]\\
    &-3m_j\operatorname{Im}(f_{LL}^{V})^*_{jk}(f_{RL}^{V})_{jk}\bigg\}.
\end{split}
\end{equation}
Finally if we define the quantity between curly brackets as $T(E)$ we have:
\begin{equation}
\label{eqn:TE}
\begin{split}
    T(E)=&\epsilon\Big[m_\ell\operatorname{Im}(f_{LL}^{V})^*_{jk}(f_{RR}^{V})_{kj}-3m_j\operatorname{Im}(f_{LL}^{V})^*_{jk}(f_{RL}^{V})_{kj}\Big]-3m_j\operatorname{Im}(f_{LL}^{V})^*_{jk}(f_{RL}^{V})_{jk},
\end{split}
\end{equation}
which is a well-known result  \cite{52}\footnote{Note that due to our Hamiltonian definition, our results are related by the exchange $f_{RL}^{V}\leftrightarrow f_{LR}^{V}$ with respect to those obtained in \cite{52}.}.\\

\begin{figure}[tbp]
    \centering
    \includegraphics{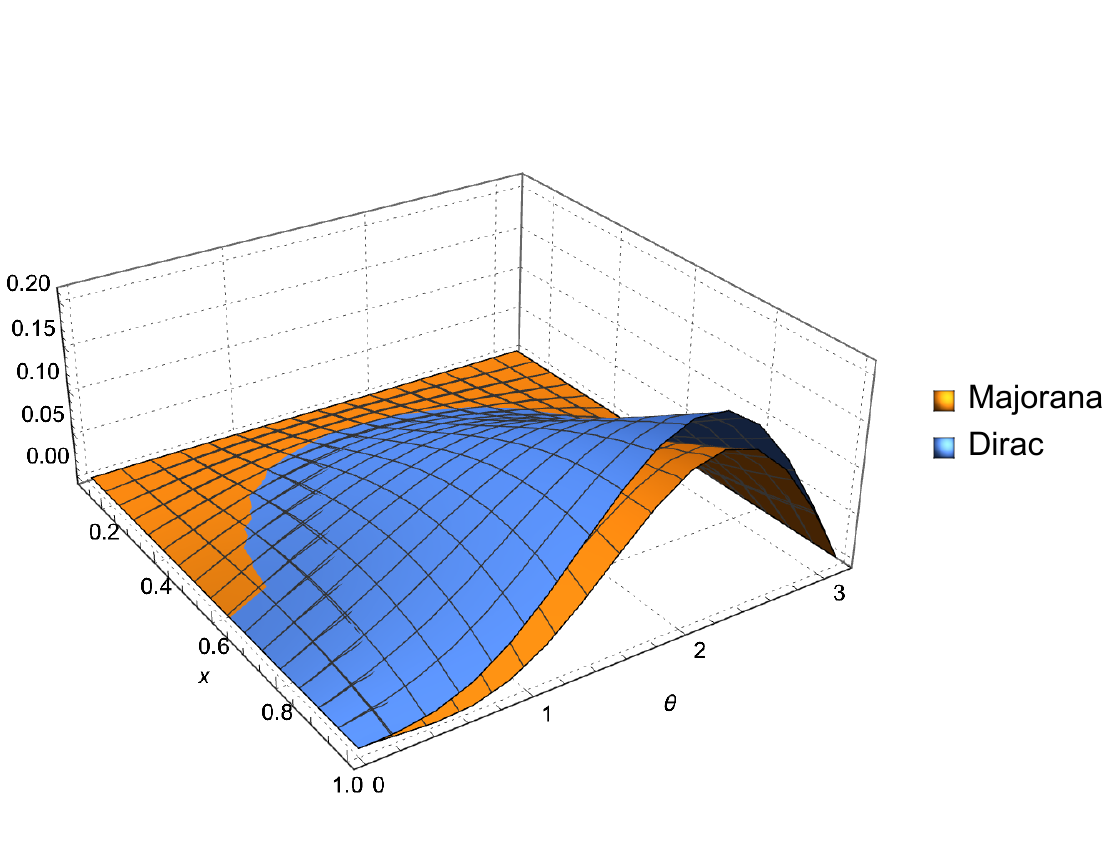}
    \caption{\label{fig: V-Adoubly}Doubly differential decay rate for the $V\pm A$ theory with $P_{T_2}$ polarization.}
\end{figure}

\begin{figure}[tbp]
    \centering
    \includegraphics{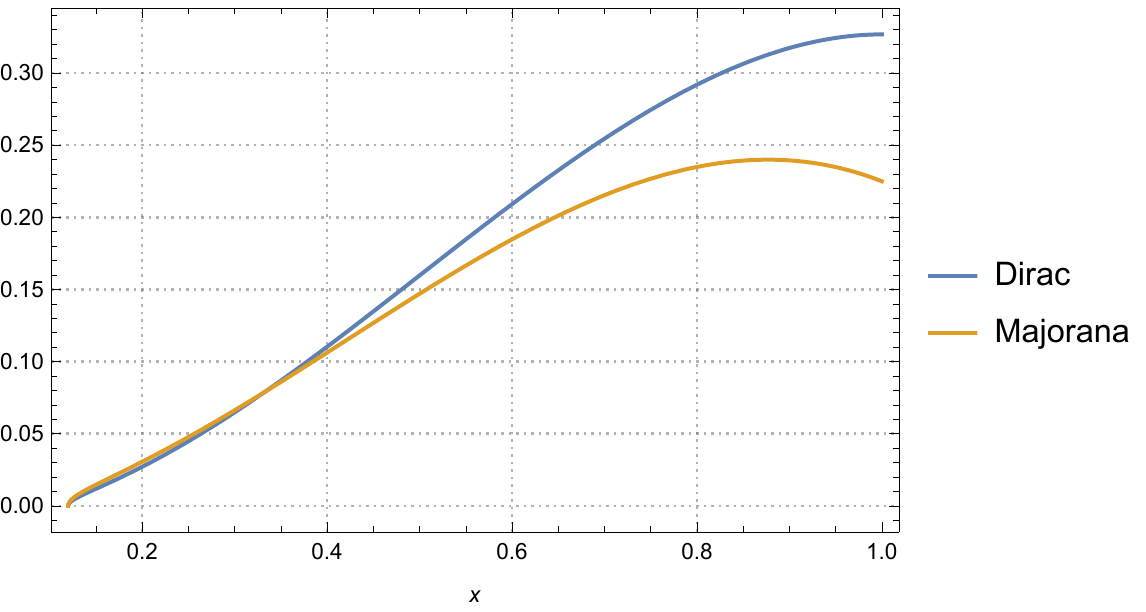}
    \caption{Energy spectrum for the $V\pm A$ theory with $P_{T_2}$ polarization.}
    \label{fig:v-aplot}
\end{figure}

\begin{figure}[tbp]
    \centering
    \includegraphics{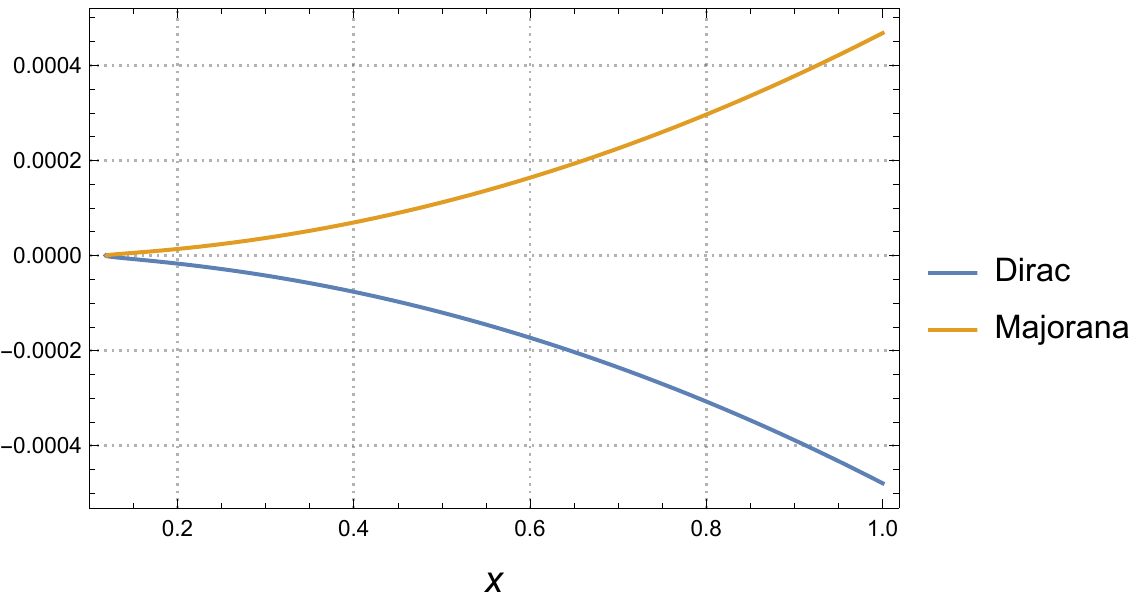}
    \caption{Neutrino mass contribution to Dirac and Majorana distributions.}
    \label{fig:DMdif}
\end{figure}

From eq.(\ref{eqn:TE}) we can see that the only Dirac contribution ($\epsilon=0$) is suppressed by a factor of neutrino mass, while the first Majorana contribution does not have a neutrino mass suppression. Thus the Majorana term dominates over the Dirac term, so that the Majorana properties of neutrinos can be observed in $T(E)$. Many other discussions and properties of these model can be found in \cite{12}. We can also plot the normalized differential decay rate in order to analyse the Dirac and Majorana differences in this model under certain conditions.\\
 If we consider a transverse-polarized final charged lepton ($F_{T_2}$), we obtained for a $\tau \rightarrow \mu \nu \Bar{\nu}$ decay:
 \begin{equation}
    \begin{split}
        \frac{d\Gamma}{dx d\cos{\theta}}=&\sum_{j,k}\frac{m_\ell}{4 \pi^3} \omega^4 G_{\ell\ell^{'}}^2\sqrt{x^2-x_0^2}\Big(F_{IS}(x)+F_{IS}'(x)-\cos{\theta}\big(F_{AS}(x)+F_{AS}'(x)\big)\Big)\\
        &+\sin{\theta}\big(F_{T_2}(x)+F_{T_2}'(x)\big).
    \end{split}
\end{equation}
Taking into account only the leading order terms and the experimental upper limits for the coupling constants, together with the suppression effects for the neutrino masses and mixings discussed before, it reduces to:
\begin{equation}
\begin{split}
&\frac{1}{N}\frac{d\Gamma}{dx d\theta }=\sin{\theta}\sqrt{x^2-x_0^2} \bigg\{x(1-x)+\frac{1}{6}(4x^2-3x-x_0^2)+0.2\epsilon x_0(1-x)-0.0005(1+\epsilon)\\
&\resizebox{\textwidth}{!}
     {$\quad\quad\left[0.1\left(x\left(1+\sqrt{1-x_0^2}\right)-x_0^2\right)-0.5\left(x_0(1-x)+x_0\sqrt{1-x_0^2} \right)\right]-\cos{\theta}\bigg[\frac{1}{3}\sqrt{x^2-x_0^2}$}\\
&\resizebox{\textwidth}{!}
     {$\quad\quad \left(1-x+\frac{1}{2}\left(4x-4+\sqrt{1-x_0^2}\right) \right)+0.0005\sqrt{x^2-x_0^2}(1+\epsilon)\left(0.5 x_0-0.1\left(1+\sqrt{1-x_0^2} \right)\right)\bigg]$}\\
&\quad\quad+\sin{\theta}\sqrt{x^2-x_0^2}\bigg[-\frac{0.2}{3}\epsilon\sqrt{1-x_0^2}+0.0005(1+\epsilon)\left(0.5\left(1+\sqrt{1-x_0^2} \right)-0.1x_0 \right)\bigg]\bigg\},
\end{split}
\end{equation}
 with $N=\frac{m_1}{4\pi^3}\omega^4 G_{\ell\ell^{'}}^2$ and $x_0<x<1$.\\
If we keep the angular dependence, the behavior of the doubly differential decay rate is shown in figure \ref{fig: V-Adoubly}.
Finally, if we integrate over all the angular dependence, the energy distribution takes the form represented in figure \ref{fig:v-aplot}.\\
\begin{figure}[tbp]
\centering
\begin{subfigure}[]{1.0\textwidth}
   \includegraphics[width=1\linewidth]{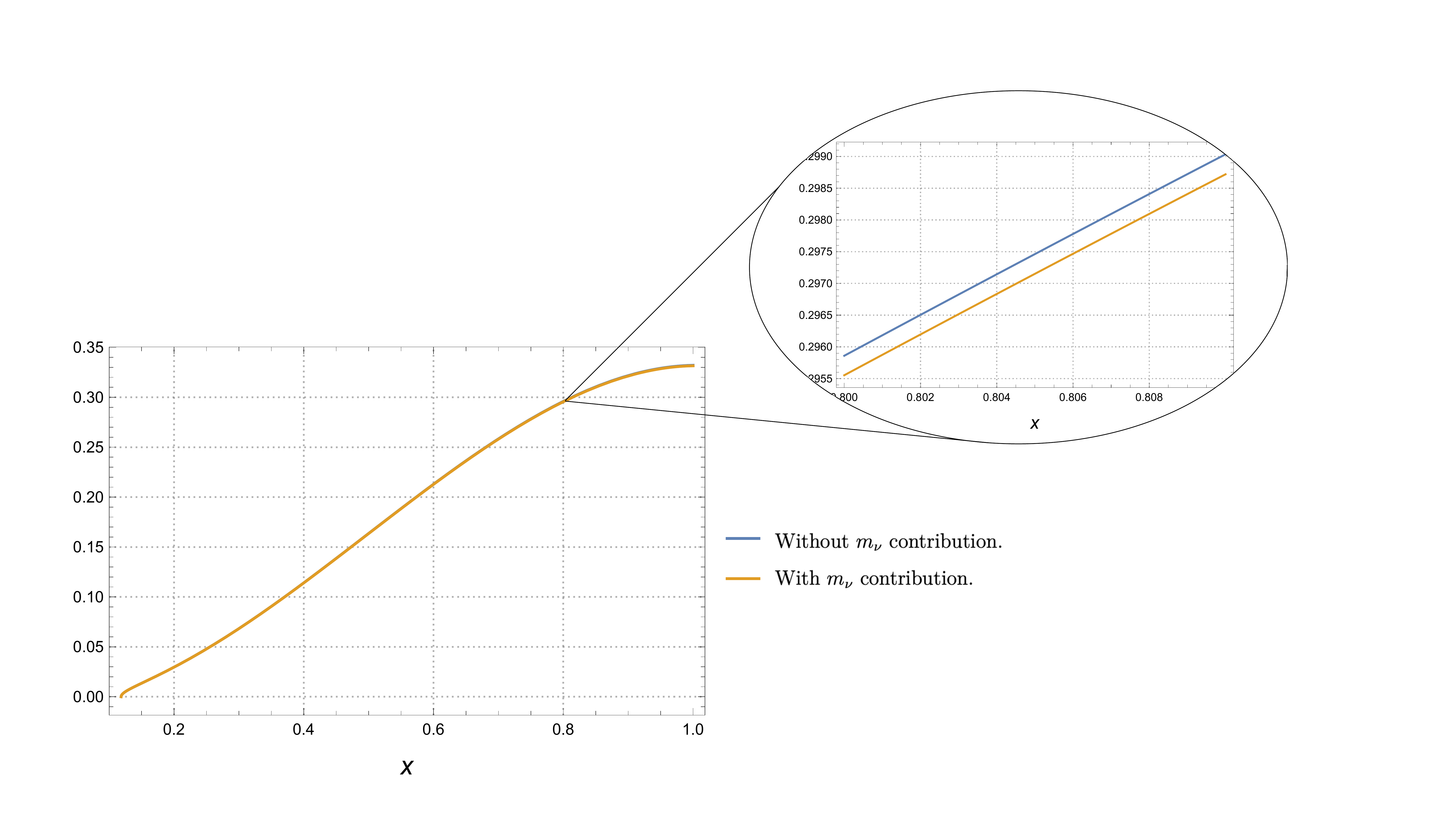}
   \caption{Dirac neutrinos.}
   \label{fig:D} 
\end{subfigure}
\begin{subfigure}[]{1.0\textwidth}
   \includegraphics[width=1\linewidth]{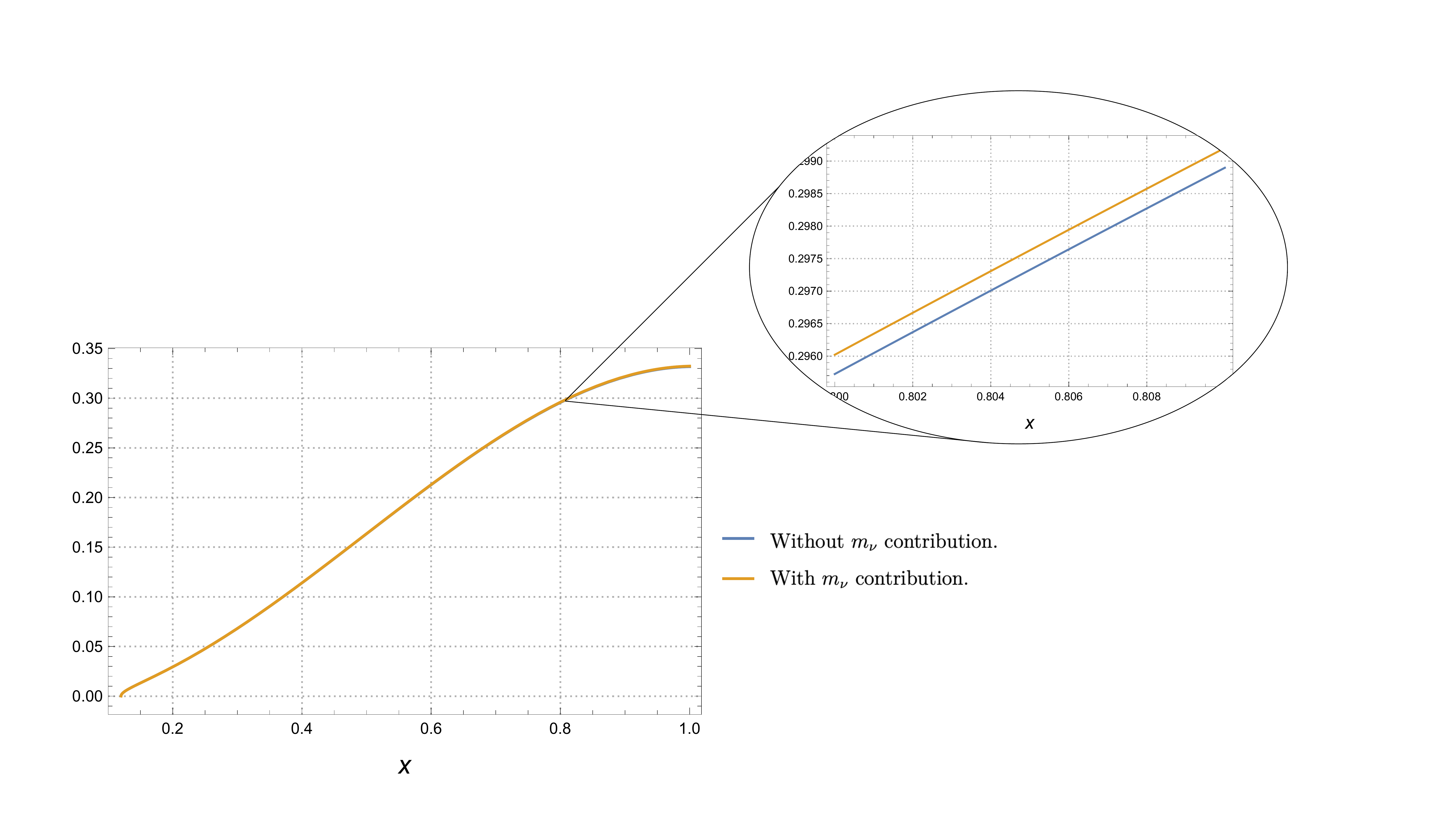}
   \caption{Majorana neutrinos.}
   \label{fig:M}
\end{subfigure}
\caption{Neutrino mass effect on the energy spectrum of the final charged-lepton.}
\label{fig:real}
\end{figure}
Thus, following the discussion, we see that the specific neutrino nature could generate measurable distortions in the energy and angular spectrum under certain circumstances. All these effects depend on the specific model used and the current limits on couplings and mixings.\\
Therefore, our model-independent result could be used to obtain model-dependent expressions in a fast way and with high theoretical precision, where all the implications of the new model-dependent physics could be discussed.\\
As a final motivation, we estimate the possible effects of the heavy neutrino sector using a realistic example. We take the $g_{LL}^V$ coupling to be dominant and consider a non-zero $g_{RR}^S$ coupling, which is one of the most motivated new physics interactions since it would couple the well-known left-handed neutrinos and it naturally appears in many beyond SM theories, it can also be tested in more detailed via the $\eta$ parameter, as we just discussed in section \ref{sec:2}. Finally, in order to obtain a non-zero contribution from the linear neutrino mass terms, we add a $g_{LR}^S$ coupling.\\
As a realistic approach, we consider the experimental mean values for the standard Michel parameters, while for the new parameters, that appear in the neutrino mass dependent terms,  we assign the following numerical values: $g_{LL}^V=0.96$, $g_{RR}^S=0.25$ and $g_{LR}^S=0.5$, which agree with the current coupling upper limits and their corresponding normalization condition.\\
Using these values, we plot the corresponding energy spectrum without any neutrino mass effect and considering the new neutrino mass terms, for Dirac and Majorana cases, which are shown in figure \ref{fig:D} and \ref{fig:M},  respectively.\\
As we just shown, the neutrino mass suppression could be of order $10^{-3}$ without any other considerations besides its mass and mixing. Now, in this realistic example we find that even considering the suppression of the specific new physics couplings and phase space structures, this neutrino mass effect could be as high as $10^{-4}$ and would lead to measurable distortions in the energy spectrum, as seen in figure \ref{fig:real}.\\
Another interesting fact in this example is that the neutrino mass terms affect in opposite way the energy distribution depending on the specific neutrino nature. For Dirac neutrinos this effect decreases the differential decay rate, while for Majorana neutrinos it increases it.\\
Actually, we can obtain the specific neutrino mass term contribution for Dirac and Majorana cases making the subtraction between the differential rate with neutrino mass effects and the one without this contribution. This result is shown in figure \ref{fig:DMdif} and it is consistent with the discussion above, a net effect of order $10^{-4}$ that is opposite for Dirac and Majorana cases.\\ 
Thus the possible distortion of the energy spectrum would not only help us to characterize the presence of a heavy neutrino sector but also to distinguish between the Dirac or Majorana nature of neutrinos.\\ 
\section{Summary and Conclusions}
\label{sec:5}
There are some physical phenomena that motivate the research of NP that is beyond the
SM. Leptonic lepton decays provide a clean laboratory where many high-precision measurements can
be made in order to test the internal consistency of the SM and reveal the signature of
possible NP.\\
In particular, we do not know whether the observed neutrinos
are Dirac or Majorana particles or if the lepton sector includes
additional fermion singlets (sterile neutrinos). The specific nature of neutrinos as well as the existence of sterile neutrinos, would affect the leptonic decays in a non-trivial way.\\
In this work we have studied the lepton decay $\ell^{-}\longrightarrow \ell^{'^{-}}{N}_j N_k$, where $N_j$ and $N_k$ are mass-eigenstate neutrinos. We have constructed its matrix element by using the most general
four-lepton effective interaction Hamiltonian and obtained the specific energy and
angular distribution of the final charged lepton, complemented with the decaying and final charged-lepton
polarization and the effects of Dirac or Majorana neutrino masses, in a way that complements all previous results, facilitating its application to model-dependent scenarios, as well as the differentiation of the possible Dirac or Majorana nature of neutrinos.\\
We have introduced generalized Michel parameters that arise due to considering finite neutrino masses and a specific neutrino nature. We write our results in a general form, classifying the Dirac and Majorana contributions with the help of a flag parameter $\epsilon=0,1$. We discuss their properties and main differences, together with some examples of its application to model-dependent theories. \\
In order to estimate the size of the neutrino mass dependent contributions we have used some of the best experimental constraints on an invisible heavy neutrino. Once the suppression of the terms has been estimated, almost all of the possible results are many orders of magnitude lower than the experimental precision of current and forthcoming experiments. Nevertheless there are some heavy masses range together with a non-negligible heavy-light mixing that lead to contributions right at the current experimental precision limit. Specifically, for the case of $\tau$-decay with one heavy final-state neutrino with a mass around $10^{2}-10^{3}$  MeV, the
linear term suppression could be of order $10^{-4}$, low enough to be measured in current and forthcoming experiments \cite{Eidelman:2015wja,Epifanov:2017kly,Belle-II:2018jsg,Bodrov:2021hfe,Lyu:2021tlb, Aiba:2021bxe}.\\
We also introduced and discussed the leading W-boson propagator correction including the final charged-lepton polarization.
Finally there is some work that can be done to complement these results. It would also be interesting to analyze other type of leptonic decays, such as radiative muon and tau decay \cite{40, radtaudecay} with Dirac and Majorana neutrinos, where new complementary information could be obtained.

\appendix
\section{Appendix: W-Boson Propagator Corrections}
\label{appendix:A}
As we already discussed, the most recent corrections induced by the $W$-boson propagator to the total and differential rates of the leptonic decay of a polarized muon and tau lepton and the numerical effect of these corrections are calculated in \cite{47} and \cite{53}.\\
Here we take their results and compare them to the standard Michel distribution form, in order to obtain the $W$-boson propagator corrections to each specific Michel parameter.\\
The energy-angle distribution of the final charged lepton in the decays  of a polarized $\mu^-$ or $\tau^-$ at rest is \cite{47}:
\begin{equation}
    \begin{split}
      \frac{d^2 \Gamma}{dx d\cos{\theta}}  =& \frac{G^2_F M^5}{192 \pi^3}\frac{x\beta}{1+\delta_{W}(M,m)}\Big\{ 3x-2x^2+r^2(3x-4)+f(x)+ r^2_{W}\big[2x^2-x^3\\
      &-2r^2(1+x-x^2+r^2)\big]-\cos{\theta}\ x\beta\big[2x-1-3r^2+g(x)+r^2_W\ x(x-2r^2)\big]\\
      & +\mathcal{O}(r^4_W)\Big\},
    \end{split}
\end{equation}
where $r_W=M/M_W$, $\beta=\frac{|\Vec{p_m}|}{E_m}=\sqrt{1-4r^2/x^2}$ and $r=m/M$. The functions $f(x)$ and $g(x)$ are the QED radiative corrections given, up to $\mathcal{O}$($\alpha$), by \cite{26}~\footnote{Higher-order corrections have been computed in refs.~\cite{vanRitbergen:1998yd,vanRitbergen:1999fi,Caola:2014daa,Fael:2020tow,Czakon:2021ybq}. See also \cite{Czarnecki:2014cxa}, on the modifications for a muon bound to a nucleus.}:\\
\begin{equation}
    \frac{2\pi}{\alpha}f(x)=(6-4x)R(x)+(6-6x)\ln{x}+\frac{1-x}{3x^2}\Big[(5+17x-34x^2)(w+\ln{x})-22x+34x^2\Big],
\end{equation}
\begin{equation}
\begin{split}
    \frac{-2\pi}{\alpha}g(x)=&(2-4x)R(x)+(2-6x)\ln{x}-\frac{1-x}{3x^2}\Big[(1+x+34x^2)(w+\ln{x})+3\\
    &-7x-32x^2+\frac{4(1-x)^2}{x}\ln{(1-x)}\Big] ,
    \end{split}
\end{equation}
where
\begin{equation}
    R(x)=2\sum_{n=1}^\infty \frac{x^n}{n^2}-\frac{1}{3}\pi^2-2+w\bigg[\frac{3}{2}+2\ln{\frac{1-x}{x}}\bigg]-\ln{x}(2\ln{x}-1)+\left(3\ln{x}-1-\frac{1}{x}\right)\ln{(1-x)},
\end{equation}
and $w=\ln(M/m)$.\\
The terms proportional to $r^2_W$ are induced by the $W$-boson propagator. Focusing only on the leading W-boson propagator contributions, we obtain:
\begin{equation}
      \frac{d^2 \Gamma}{dx d\cos{\theta}}  = \frac{G^2_F M^5}{192 \pi^3}\frac{x}{1+3/5(M/M_W)^2}\bigg\{ 3x-2x^2+ r^2_{W}\big[2x^2-x^3\big]-\cos{\theta}\ x\big[2x-1+r^2_W\ x^2\big]\bigg\}.
\end{equation}
So that the Michel parameter, defined in terms of the energy spectrum (neglecting the $x_0$ dependence):
\begin{equation}
    \frac{d\Gamma_{\ell\rightarrow \ell^{'}} }{dxd\cos{\theta}}=\frac{G_{\ell\ell^{'}}^2 M^5 }{192\pi^3}x\left\{6x(1-x)+\frac{4}{3}\rho\left(4x^2-3x\right)-2\xi x\cos{\theta}\left(1-x+\frac{2}{3}\delta(4x-3)\right)\right\},
\end{equation}
is given by:
\begin{equation}
    \begin{split}
       &\bullet\rho\rightarrow\rho_{eff}=\frac{3}{4}+\frac{3}{2}\left(\frac{M}{M_W}\right)^2,
    \end{split}
\end{equation}
where $G_{\ell\ell^{'}}^2=G^2_F/(1+\delta_{W}(m_{\ell},m_{\ell^{'}}))$ and the remaining parameters have no further modification.\\
Finally, as far as we know, the W-boson corrections in the case of a final charged-lepton polarization have not been computed yet. Here, we present their expression at first order in $r_W$ and neglecting the $x_0$ dependence.\\
The final-lepton polarization contribution of the differential decay rate is then:
\begin{equation}
\begin{split}
    \frac{d\Gamma_{\ell\rightarrow \ell^{'}}}{dxd\cos{\theta}}=&\frac{G_{\ell\ell^{'}}^2 M^5 }{64\pi^3}x\bigg\{-\left[\frac{1}{6}x(-2x+3)+r_W^2\frac{1}{6}x(2x-x^2)\right]\cos{\phi}\\
    &+\frac{1}{6}\bigg[(2x^2-x)+r_W^2 x^3\bigg]\cos{\theta}\cos{\phi}\bigg\}.
    \end{split}
\end{equation}
So that the Michel parameters, defined in terms of the energy spectrum (neglecting the $x_0$ dependence):
\begin{equation}
\begin{split}
    \frac{d\Gamma_{\ell\rightarrow \ell^{'}}}{dxd\cos{\theta}}&=\frac{G_{\ell\ell^{'}}^2 M^5 }{64\pi^3}x\left\{-F_{IP}\cos{\phi}+F_{AP}\cos{\theta}\cos{\phi}\right\}\\
    &=\frac{G_{\ell\ell^{'}}^2 M^5 }{64\pi^3}x\bigg\{-\frac{1}{54}x\left[9\xi^{'}(3-2x)+4\xi(\delta-\frac{3}{4})(4x-3) \right]\cos{\phi}+\frac{1}{6}\bigg[\xi^{''}(2x^2-x)\\
    &+4(\rho-\frac{3}{4})(4x^2-3x) \bigg]\cos{\theta}\cos{\phi}\bigg\},
\end{split}
\end{equation}
are given by:
\begin{equation}
    \begin{split}
       &\bullet\xi^{'}\rightarrow\xi^{'}_{eff}=1+\left(\frac{M}{M_W}\right)^2\\
       &\bullet\xi\Big(\delta-\frac{3}{4}\Big)\rightarrow\xi\Big(\delta-\frac{3}{4}\Big)_{eff}=0+\frac{9}{4}\left(\frac{M}{M_W}\right)^2
    \end{split}
\end{equation}
and the remaining ones are not modified.\\
However, it is interesting to emphasize that even if the $\rho$, $\xi$ and $\xi\delta$ parameters appear in both, the initial and final lepton polarization contribution, they are modified in an independent way in each of these contributions. Then, we can not simple redefine each Michel parameter with an effective and global modification, since we need to specify where they appear in the differential decay rate.\\
In order to keep the least number of parameters in our expressions, we suggest not to redefine new effective parameters and just take into account independently the W-boson propagator corrections as well as the QED contributions \cite{47} to the differential decay rate instead.

\section{Appendix: Fierz Transformations}
\label{appendix:B}
A Fierz transformation is an operation to change the order of the fermion fields in the four-fermion interaction Lagrangian. Two Hamiltonians related by a Fierz transformation are physically equivalent. This is useful, since the implementation of an specific form of the Hamiltonian can make the calculations simpler, the physical interpretation more intuitive, etc.\\
In this appendix we summarize the specific Fierz transformations between the most commonly used set of Hamiltonians in lepton decay. Our main goal is to show the Fierz transformation between the Hamiltonian used in \cite{51} and the one employed for this work, in order to prove the consistency of both results. Further information about Fierz identities and their specific derivation can be seen in \cite{55,56}.\\
The leptonic decays can be described by the most general four-fermion contact interaction Hamiltonian. The contact interaction allows the use of equivalent Hamiltonians, which differ in the way the fermions are grouped together.\\
As discussed in \cite{57}, the older literature preferred a $charge$ $retention$ form (CRF) with parity-odd and parity-even terms in which the charged-leptons, as the usually detected particles, were grouped together. The CRF Hamiltonian takes the following form:
\begin{equation}
\label{eqn:CRFH}
    \mathcal{H}= \frac{G_{ll^{'}}}{\sqrt{2}}\sum_i\bigg\{\big(\overline{f}_1 \Gamma_i f_4 \big) \Big[D_i(\overline{f}_3 \Gamma^i f_2 \big)+
    D_i^{'} (\overline{f}_3 \Gamma^{i}\gamma^5 f_2 \big)\Big]+ h.c. \bigg\},
\end{equation}
with i= S, P, A, V, T and the Lorentz bilinears are defined as:\footnote{Note, in particular, the factor i in the definition of $\Gamma_P$.}
\begin{equation}
\label{eqn:FH1}
    \begin{split}
        &\Gamma_S\equiv\mathbb{I}, \quad\quad\quad \Gamma_P\equiv i\gamma^5,\quad\quad\quad \Gamma_V\equiv\gamma^\mu,\\
        &\Gamma_A\equiv \gamma^{\mu}\gamma^5, \quad\quad\quad \Gamma_T\equiv \frac{1}{\sqrt{2}}\sigma^{\mu\nu},
    \end{split}
\end{equation}
with the identification $f_1=\ell^{'}$, $f_2=\nu_{\ell^{'}}$, $f_3=\nu_{\ell}$ and $f_4=\ell$.\\
This had the advantage that limits to some coupling constants could be obtained from then existing results, but is not adapted to represent a charged boson exchange.\\
For this purpose, it is used a $charge$ $changing$ form (CCF), where the charged leptons are grouped with their neutrinos\footnote{For this reason also called flavor retention form.} and which is adapted to charged boson exchange and results in absolute values of differences of coupling constants. This CCF Hamiltonian takes the form:
\begin{equation}
\label{eqn:FH2}
    \mathcal{H}= \frac{G_{ll^{'}}}{\sqrt{2}}\sum_i\bigg\{\big(\overline{f}_1 \Gamma_i f_2 \big) \Big[C_i(\overline{f}_3 \Gamma^i f_4 \big)+
    C_i^{'} (\overline{f}_3 \Gamma^{i}\gamma^5 f_4 \big)\Big]+ h.c. \bigg\}
\end{equation}
However both of these forms are complicated by the fact that a fully parity-violating interaction, such as e.g. the V-A interaction, is represented by four coupling constants $C_V$ , $C^{'}_V$ , $C_A$ and $C^{'}_A$.\\
Finally there is the $helicity$ $projection$ form (HPF) Hamiltonian, which is the most used nowadays, where the canonical V-A interaction is particularly simple because then $g_{LL}^V=1$ while all other coupling constants vanish and the contributions from new physics are immediately identified. Its explicit form, used in this work, is:
\begin{equation}
\label{eqn:FH3}
\begin{split}
     \mathcal{H}&=\frac{G_{\ell\ell^{'}}}{\sqrt{2}}\bigg\{ g_{LL}^S\left[\Bar{\ell}^{'} (1+\gamma^5) \nu_{{\ell^{'}}}\right]\Big[\Bar{\nu}_{\ell}(1-\gamma^5) \ell\Big]+g_{LL}^V\left[\Bar{\ell}^{'} \gamma^\mu (1-\gamma^5) \nu_{{\ell^{'}}}\right]\Big[\Bar{\nu}_{\ell}\gamma_\mu (1-\gamma^5) \ell\Big]\\
     &+g_{RR}^S\left[\Bar{\ell}^{'}(1-\gamma^5) \nu_{{\ell^{'}}}\right]\Big[\Bar{\nu}_{\ell}(1+\gamma^5) \ell\Big]+g_{RR}^V\left[\Bar{\ell}^{'} \gamma^\mu (1+\gamma^5) \nu_{{\ell^{'}}}\right]\Big[\Bar{\nu}_{\ell}\gamma_\mu (1+\gamma^5) \ell\Big]\\
     &+g_{LR}^S\left[\Bar{\ell}^{'}(1+\gamma^5) \nu_{{\ell^{'}}}\right]\Big[\Bar{\nu}_{\ell}(1+\gamma^5) \ell\Big]+g_{LR}^V\left[\Bar{\ell}^{'} \gamma^\mu (1-\gamma^5) \nu_{{\ell^{'}}}\right]\Big[\Bar{\nu}_{\ell}\gamma_\mu (1+\gamma^5) \ell\Big]\\
     &+g_{LR}^T\left[\Bar{\ell}^{'} \frac{\sigma^{\mu\nu}}{\sqrt{2}}(1+\gamma^5) \nu_{{\ell^{'}}}\right]\Big[\Bar{\nu}_{\ell}\frac{\sigma_{\mu\nu}}{\sqrt{2}}(1+\gamma^5) \ell\Big]+g_{RL}^S\left[\Bar{\ell}^{'}(1-\gamma^5) \nu_{{\ell^{'}}}\right]\Big[\Bar{\nu}_{\ell}(1-\gamma^5) \ell\Big]\\
     &+g_{RL}^V\left[\Bar{\ell}^{'} \gamma^\mu (1+\gamma^5) \nu_{{\ell^{'}}}\right]\Big[\Bar{\nu}_{\ell}\gamma_\mu (1-\gamma^5)\ell\Big]+ g_{RL}^T\left[\Bar{\ell}^{'} \frac{\sigma^{\mu\nu}}{\sqrt{2}}(1-\gamma^5) \nu_{{\ell^{'}}}\right]\Big[\Bar{\nu}_{\ell}\frac{\sigma_{\mu\nu}}{\sqrt{2}}(1-\gamma^5) \ell\Big]\bigg\}.
     \end{split}
\end{equation}
 Thus, we can obtain the specific Fierz transformation that relates each pair of Hamiltonians.\\
The Fierz transformation that relates the CRF and CCF Hamiltonians is
\cite{58}:
\begin{equation}
\begin{pmatrix}
D_S\\
D_P \\
D_V \\
D_A\\
D_T
\end{pmatrix}
=\frac{1}{4}
\begin{pmatrix}
-1 & 1 & -4 & 4 & -6\\
1 & -1 & -4 & 4 & 6\\
-1 & -1 & 2 & 2 & 0\\
1 & 1 & 2 & 2 & 0\\
-1 & 1 & 0 & 0 & 2
\end{pmatrix}
\begin{pmatrix}
C_S \\
C_P \\
C_V \\
C_A\\
C_T
\end{pmatrix}
\end{equation}
\begin{equation}
\begin{pmatrix}
D_S^{'}\\
D_P^{'} \\
D_V^{'} \\
D_A^{'}\\
D_T^{'}
\end{pmatrix}
=\frac{1}{4}
\begin{pmatrix}
-1 & 1 & 4 & -4 & -6\\
1 & -1 & 4 & -4 & 6\\
1 & 1 & 2 & 2 & 0\\
-1 & -1 & 2 & 2 & 0\\
-1 & 1 & 0 & 0 & 2
\end{pmatrix}
\begin{pmatrix}
C_S^{'} \\
C_P^{'} \\
C_V^{'} \\
C_A^{'}\\
C_T^{'}
\end{pmatrix}
\end{equation}
In the same way, the Fierz transformation that relates the CCF and HPF Hamiltonians is \cite{57}:
\begin{equation*}
    \begin{matrix}
C_S\\
C_P
\end{matrix} 
\Bigg\}= \big(g^{S}_{LL}+g^{S}_{RR} \big)\pm \big(g^{S}_{LR}+g^{S}_{RL} \big)
\end{equation*}
\begin{equation*}
\hspace*{0.3cm}
    \begin{matrix}
C_S^{'}\\
C_P{'}
\end{matrix} 
\Bigg\}=-\big(g^{S}_{LL}-g^{S}_{RR} \big)\pm \big(g^{S}_{LR}-g^{S}_{RL} \big)
\end{equation*}
\begin{equation*}
    \begin{matrix}
C_V\\
C_A
\end{matrix} 
\Bigg\}=\big(g^{V}_{RR}+g^{V}_{LL} \big)\pm \big(g^{V}_{RL}+g^{V}_{LR} \big)
\end{equation*}
\begin{equation*}
    \begin{matrix}
C_V{'}\\
C_A{'}
\end{matrix} 
\Bigg\}=\big(g^{V}_{RR}-g^{V}_{LL} \big)\mp \big(g^{V}_{RL}-g^{V}_{LR} \big)
\end{equation*}
\begin{equation}
\hspace*{-2.5cm}
    \begin{matrix}
C_T\\
C_T{'}
\end{matrix} 
\Bigg\}=2\big(g^{T}_{LR}\pm g^{T}_{RL} \big)
\end{equation}

Thus, the final relation between the CRF and HPF is given by:
\begin{equation}
\begin{pmatrix}
D_S\\
D_P \\
D_V \\
D_A\\
D_T
\end{pmatrix}
=\frac{1}{4}
\begin{pmatrix}
-1 & 1 & -4 & 4 & -6\\
1 & -1 & -4 & 4 & 6\\
-1 & -1 & 2 & 2 & 0\\
1 & 1 & 2 & 2 & 0\\
-1 & 1 & 0 & 0 & 2
\end{pmatrix}
\begin{pmatrix}
g^{S}_{LL}+g^{S}_{RR}+g^{S}_{LR}+g^{S}_{RL}\\
g^{S}_{LL}+g^{S}_{RR}-g^{S}_{LR}-g^{S}_{RL}\\
g^{V}_{RR}+g^{V}_{LL}+g^{V}_{RL}+g^{V}_{LR}\\
g^{V}_{RR}+g^{V}_{LL}-g^{V}_{RL}-g^{V}_{LR}\\
2\big(g^{T}_{LR}+g^{T}_{RL} \big)
\end{pmatrix}
\end{equation}
\begin{equation}
\begin{pmatrix}
D_S^{'}\\
D_P^{'} \\
D_V^{'} \\
D_A^{'}\\
D_T^{'}
\end{pmatrix}
=\frac{1}{4}
\begin{pmatrix}
-1 & 1 & 4 & -4 & -6\\
1 & -1 & 4 & -4 & 6\\
1 & 1 & 2 & 2 & 0\\
-1 & -1 & 2 & 2 & 0\\
-1 & 1 & 0 & 0 & 2
\end{pmatrix}
\begin{pmatrix}
-g^{S}_{LL}+g^{S}_{RR}+g^{S}_{LR}-g^{S}_{RL}\\
-g^{S}_{LL}+g^{S}_{RR}-g^{S}_{LR}+g^{S}_{RL}\\
g^{V}_{RR}-g^{V}_{LL}-g^{V}_{RL}+g^{V}_{LR}\\
g^{V}_{RR}-g^{V}_{LL}+g^{V}_{RL}-g^{V}_{LR}\\
2\big(g^{T}_{LR}-g^{T}_{RL} \big)
\end{pmatrix}
\end{equation}
Finally, it is important to note that the CRF Hamiltonian used in \cite{51} differs from eq.(\ref{eqn:CRFH}) in the definition of $\Gamma_P$. In Shrock's paper $\Gamma_P=\gamma^5$, this change will induce a minus sign in the definition of the constants $C_P$ and $C^{'}_P$ previously introduced. Thus, the Fierz transformation between the Shrocks's Hamiltonian (using his couplings conventions) and the Hamiltonian used in this work is:
\begin{equation}
\begin{pmatrix}
g_S\\
g_P \\
g_V \\
g_A\\
g_T
\end{pmatrix}
=\frac{1}{4}
\begin{pmatrix}
-1 & 1 & -4 & 4 & -6\\
-1 & 1 & 4 & -4 & -6\\
-1 & -1 & 2 & 2 & 0\\
1 & 1 & 2 & 2 & 0\\
-1 & 1 & 0 & 0 & 2
\end{pmatrix}
\begin{pmatrix}
g^{S}_{LL}+g^{S}_{RR}+g^{S}_{LR}+g^{S}_{RL}\\
g^{S}_{LL}+g^{S}_{RR}-g^{S}_{LR}-g^{S}_{RL}\\
g^{V}_{RR}+g^{V}_{LL}+g^{V}_{RL}+g^{V}_{LR}\\
g^{V}_{RR}+g^{V}_{LL}-g^{V}_{RL}-g^{V}_{LR}\\
2\big(g^{T}_{LR}+g^{T}_{RL} \big)
\end{pmatrix}
\end{equation}
\begin{equation}
\begin{pmatrix}
g_S^{'}\\
g_P^{'} \\
g_V^{'} \\
g_A^{'}\\
g_T^{'}
\end{pmatrix}
=\frac{1}{4}
\begin{pmatrix}
-1 & 1 & 4 & -4 & -6\\
-1 & 1 & -4 & 4 & -6\\
1 & 1 & 2 & 2 & 0\\
-1 & -1 & 2 & 2 & 0\\
-1 & 1 & 0 & 0 & 2
\end{pmatrix}
\begin{pmatrix}
-g^{S}_{LL}+g^{S}_{RR}+g^{S}_{LR}-g^{S}_{RL}\\
-g^{S}_{LL}+g^{S}_{RR}-g^{S}_{LR}+g^{S}_{RL}\\
g^{V}_{RR}-g^{V}_{LL}-g^{V}_{RL}+g^{V}_{LR}\\
g^{V}_{RR}-g^{V}_{LL}+g^{V}_{RL}-g^{V}_{LR}\\
2\big(g^{T}_{LR}-g^{T}_{RL} \big)
\end{pmatrix}
\end{equation}
After some work of simplification, the specific relations between constants are: 
\begin{equation}
    \begin{split}
&g_S=-\frac{1}{2}g^{S}_{LR}-\frac{1}{2}g^{S}_{RL}-2g^{V}_{RL}-2g^{V}_{LR}-3g^{T}_{LR}-3g^{T}_{RL}\\
&g_S^{'}=-\frac{1}{2}g^{S}_{LR}+\frac{1}{2}g^{S}_{RL}-2g^{V}_{RL}+2g^{V}_{LR}-3g^{T}_{LR}+3g^{T}_{RL}\\
&g_P=-\frac{1}{2}g^{S}_{LR}-\frac{1}{2}g^{S}_{RL}+2g^{V}_{RL}+2g^{V}_{LR}-3g^{T}_{LR}-3g^{T}_{RL}\\
&g_P^{'}=-\frac{1}{2}g^{S}_{LR}+\frac{1}{2}g^{S}_{RL}+2g^{V}_{RL}-2g^{V}_{LR}-3g^{T}_{LR}+3g^{T}_{RL} \\
&g_V=g^{V}_{RR}+g^{V}_{LL}-\frac{1}{2}g^{S}_{LL}-\frac{1}{2}g^{S}_{RR}\\
&g_V^{'}=g^{V}_{RR}-g^{V}_{LL}-\frac{1}{2}g^{S}_{LL}+\frac{1}{2}g^{S}_{RR} \\
&g_A=g^{V}_{RR}+g^{V}_{LL}+\frac{1}{2}g^{S}_{LL}+\frac{1}{2}g^{S}_{RR}\\
&g_A^{'}=g^{V}_{RR}-g^{V}_{LL}+\frac{1}{2}g^{S}_{LL}-\frac{1}{2}g^{S}_{RR}\\
&g_T=-\frac{1}{2}g^{S}_{LR}-\frac{1}{2}g^{S}_{RL}+g^{T}_{LR}+g^{T}_{RL}\\
&g_T^{'}=-\frac{1}{2}g^{S}_{LR}+\frac{1}{2}g^{S}_{RL}+g^{T}_{LR}-g^{T}_{RL}
    \end{split}
\end{equation}
These relations helped us to verify the consistency of our results with the previous ones.

\acknowledgments
We express our gratitude to Denis Epifanov for emphasizing to us the timeliness of this study, which is fostered by the conscientious work of him and other brilliant experimentalists. J. M. is indebted to Conacyt funding his Ms. Sc. and Ph. D. The work of G. L. C. was supported by Ciencia de Frontera Conacyt (M\'exico) project No. 428218. P. R. acknowledges the financial support of Cátedras Marcos Moshinsky (Fundación Marcos
Moshinsky) and Conacyt’s project within ‘Paradigmas y Controversias de la Ciencia 2022’,
number 319395.



\begin{thebibliography}{99}

\bibitem{1}
Particle Data Group, 
\emph{Review of Particle Physics},  PTEP 2022 (2022) 083C01.

\bibitem{2}
Kai Schmidt-Hoberg, \emph{Phenomenology of Physics beyond the Standard Model} (2015).

\bibitem{3}
B. Pontecorvo, \emph{Mesonium and anti-Mesonium}, Zh. Eksp. Teor. Fiz. {\bf 34} (1957) 247.

\bibitem{4}
Z. Maki, M. Nakagawa, and S. Sakata, \emph{Remarks on the unified model of elementary particles}, Prog. Theor. Phys.{\bf 28} (1962) 870.

\bibitem{5}
Anupama Atre, Tao Han, Silvia Pascoli, Bin Zhang, \emph{The Search for Heavy Majorana Neutrinos} (2009),
JHEP {\bf 05} (2009) 030
.

\bibitem{6}
R. N. Mohapatra and J. W. F. Valle, \emph{Neutrino Mass and Baryon Number Nonconservation in Superstring Models}, Phys. Rev. D {\bf 34}, (1986) 1642.

\bibitem{7}
J. Bernabéu, A. Santamaría, J. Vidal, A. Méndez, and J. W. F. Valle, \emph{Lepton Flavor Nonconservation at High-Energies in a Superstring Inspired Standard Model}, Phys. Lett. B {\bf 187} (1987) 303.

\bibitem{8}
M. Malinsky, J.C. Romao, and J. W.F. Valle, \emph{Novel supersymmetric SO(10) seesaw mechanism}, Phys. Rev. Lett. {\bf 95} (2005) 161801.

\bibitem{9}
A. G. Dias , C. A. de S. Pires , P. S. Rodrigues da Silva, A. Sampieri, \emph{A Simple Realization of the Inverse Seesaw Mechanism} (2012),
arxiv:1206.2590.

\bibitem{10}
G. Hernández-Tomé, J.I. Illana , M. Masip, G. López Castro and P. Roig, \emph{Effects of heavy Majorana neutrinos on lepton flavor violating processes}, Phys. Rev. D {\bf 101} (2020) 075020 

\bibitem{11}
Takeshi Fukuyama and Joji Tsumura, \emph{Detecting Majorana nature of neutrinos in muon and tau decay} (2008),
arxiv:0809.5221.

\bibitem{12}
Masaru Doi, Tsuneyuki Kotani, Hiroyuki Nishiura, \emph{New Parametrization in Muon Decay and the Type of Emitted Neutrino} (2005), 
arxiv:hep-ph/0502136.

\bibitem{13}
Duarte, L., \textit{et al.}, \emph{Majorana neutrinos with effective interactions in B decays}, Eur. Phys. J. C {\bf 79} (2019) 593.

\bibitem{14}
S.N. Gninenko. \emph{The MiniBooNE anomaly and heavy neutrino decay} (2009),
arxiv:0902.3802.

\bibitem{15}
A.Pich, \emph{Precision Tau Physics} (2013),
Prog.Part.Nucl.Phys. {\bf 75} (2014) 41-85.

\bibitem{16}
A.Pich, \emph{Precision tests of the Standard Model} (1997),
arxiv:hep-ph/9711279.

\bibitem{17}
T.P. Gorringe, D.W. Hertzog, \emph{Precision Muon Physics} (2015),
Prog.Part.Nucl.Phys. {\bf 84} (2015) 73-123.

\bibitem{18}
Y. Amhis, \textit{et al.} (HFLAV Collaboration), \emph{Averages of $b$-hadron, $c$-hadron, and $\tau$-lepton properties as of summer 2016}, Eur. Phys. J. C {\bf 77} (2017) 895.

\bibitem{Muong-2:2006rrc}
G.~W.~Bennett \textit{et al.} [Muon g-2],
\emph{Final Report of the Muon E821 Anomalous Magnetic Moment Measurement at BNL},
Phys. Rev. D \textbf{73} (2006), 072003
.

\bibitem{Muong-2:2021ojo}
B.~Abi \textit{et al.} [Muon g-2],
\emph{Measurement of the Positive Muon Anomalous Magnetic Moment to 0.46 ppm},
Phys. Rev. Lett. \textbf{126} (2021) no.14, 141801
.

\bibitem{Aoyama:2020ynm}
T.~Aoyama, N.~Asmussen, M.~Benayoun, J.~Bijnens, T.~Blum, M.~Bruno, I.~Caprini, C.~M.~Carloni Calame, M.~C\`e and G.~Colangelo, \textit{et al.}
\emph{The anomalous magnetic moment of the muon in the Standard Model},
Phys. Rept. \textbf{887} (2020), 1-166
.



\bibitem{19}
 L. Michel., \emph{Interaction between four half spin particles and the decay of the $\mu$ meson}, Proc. Phys. Soc. {\bf A 63} (1950) 514, 1371.

\bibitem{Bouchiat:1957zz}
C.~Bouchiat and L.~Michel,
\emph{Theory of $\mu$-Meson Decay with the Hypothesis of Nonconservation of Parity},
Phys. Rev. \textbf{106} (1957), 170-172.

\bibitem{20}
W. Fetscher, H.J. Gerber and K.F. Johnson, \emph{Muon Decay: Complete Determination of the Interaction and Comparison with the Standard Model}, Phys. Lett. {\bf B173} (1986) 102.

\bibitem{21}
W. Fetscher, \emph{Helicity dependence of the electron-neutrino energy spectrum from the decay of unpolarized muons}, Phys. Rev. D {\bf 49} (1994) 5945.


\bibitem{23}
C.A. Gagliardi, R.E. Tribble, and N.J. Williams. \emph{Global analysis of muon decay measurements} (2005),
arxiv:hep-ph/0509069.

\bibitem{24}
 R. E. Behrends, R. J. Finkelstein, and A. Sirlin. \emph{Radiative corrections to decay
processes}, Phys. Rev. Lett. {\bf101} (1956) 866–873.

\bibitem{25}
S. M. Berman. \emph{Radiative corrections to muon and neutron decay}, Phys. Rev.
{\bf112} (1958) 267–270.

\bibitem{26}
Toichiro Kinoshita and Alberto Sirlin. \emph{Radiative corrections to fermi interactions},
Phys. Rev. {\bf113} (1959) 1652–1660.

\bibitem{27}
Stahl, A. \emph{The Michel parameter $\eta$ in $\tau$ decays}, Phys. Lett. {\bf324} (1994) 121–124.


\bibitem{28}
SLD Collaboration, \emph{Measurement of the tau-neutrino helicity and the Michel parameters in polarized $e^+ e^-$ collisions}, Phys. Rev. Lett. {\bf78} (1997) 4691.

\bibitem{29}
ALEPH Collaboration, \emph{Measurement of the Michel parameters and the nu/tau helicity in tau lepton decays}, Eur. Phys. J. C {\bf22} (2001) 217.

\bibitem{30}
DELPHI Collaboration, \emph{A Study of the Lorentz structure in tau decays}, Eur. Phys. J. C {\bf16} (2000) 229.

\bibitem{31}
OPAL Collaboration, \emph{Measurement of the Michel parameters in leptonic tau decays} , Eur. Phys. J. C {\bf 8} (1999) 3.

\bibitem{32}
L3 Collaboration, \emph{Measurement of the Michel parameters and the average tau-neutrino helicity from tau decays at LEP}, Phys. Lett. B {\bf438} (1998) 405.

\bibitem{33}
ARGUS Collaboration, \emph{Determination of the Michel parameters rho, xi and delta in tau lepton decays with tau $\longrightarrow$ rho neutrino tags},  Phys. Lett. B {\bf431} (1998) 179.

\bibitem{34}
CLEO Collaboration, \emph{Determination of the Michel parameters and the tau-neutrino helicity in tau decay}, Phys. Lett. D {\bf56} (1997) 5320.

\bibitem{35}
TWIST Collaboration, \emph{Precision muon decay measurements and improved constraints on the weak interaction}, Phys. Rev. D {\bf85} (2012) 092013.

\bibitem{36}
N. Danneberg \textit{et al.}, \emph{Muon decay: Measurement of the transverse polarization of the decay positrons and its implications for the Fermi coupling constant and time reversal invariance}, Phys. Rev. Lett. {\bf94} (2005) 021802.

\bibitem{37}
B. Balke \textit{et al.}, \emph{Precise Measurement of the Asymmetry Parameter Delta in Muon Decay}, Phys. Rev. D {\bf37} (1988) 587.

\bibitem{38}
I. Beltrami \textit{et al.}, \emph{Muon Decay: Measurement of the Integral Asymmetry Parameter}, Phys. Lett. B {\bf194} (1987) 326.

\bibitem{radtaudecay}
N. Shimizu et al.(Belle), \emph{Measurement of the $\tau$ Michel parameters $\bar{\eta}$ and $\xi\kappa$ in the radiative leptonic decay $\tau^{-} \rightarrow \ell^{-}\nu_\tau \bar{\nu}_\ell \gamma$}, Prog. Theo. Exp. Phys. {\bf2}, (2018) 023C01.

\bibitem{stahl}
A. Stahl and H. Voss, \emph{Testing the Lorentz structure of the charged weak current in $\tau$-decays}, Z. Phys. C {\bf74} (1997) 73.

\bibitem{39}
A. Flores-Tlalpa, G. L\'opez Castro and P. Roig , \emph{Five-body leptonic decays of muon and tau leptons}, JHEP {\bf04} (2016) 185.

\bibitem{40}
A.B. Arbuzov, T.V. Kopylova, \emph{Michel parameters in radiative muon decay}, JHEP {\bf09} (2016) 109.

\bibitem{41}
Sasaki Junya and Belle Collaboration, \emph{Study of five-body leptonic decays of tau at Belle experiment}, J. Phys. Conf. Ser. {\bf912} (2017) 012002.

\bibitem{42}
P. Langacker and D. London, \emph{Analysis of Muon Decay With Lepton Number Nonconserving Interactions}, Phys. Rev.D {\bf39} (1989) 266.

\bibitem{43}
M. M. Giannini, \emph{Muon decay and neutrino mass} (2022), 
arxiv:2207.02717

\bibitem{Giannini:2022ilc}
M.~M.~Giannini,
\emph{An improved upper limit for the (muon based) neutrino mass},
[arXiv:2207.02718 [hep-ph]].

\bibitem{51}
Robert E. Shrock, \emph{Pure Leptonic Decays With Massive Neutrinos And Arbitrary Lorentz Structure}, Phys. Lett. B {\bf112} (1982) 382.

\bibitem{44}
Ansgar Denner, H. Eck, O. Hahn, and J. Kublbeck, \emph{Feynman rules for fermion number violating interactions}, Nucl. Phys. B {\bf387} (1992) 467-481.

\bibitem{52}
M.Doi, T.Kotani and E.Takasugi, \emph{Double Beta Decay and Majorana Neutrinos}, Progress of Theoretical Physics Supplement, Volume 83 (1985).

\bibitem{45}
André de Gouvea, Andrew Kobach, \emph{Global Constraints on a Heavy Neutrino} (2016),     Phys.Rev.D 93 (2016) 3, 033005.

\bibitem{46}
Andrew Kobach and Sean Dobbs, \emph{Heavy Neutrinos and the Kinematics of Tau Decays} (2015), 
Phys.Rev.D 91 (2015) 5, 053006.

\bibitem{BaBar:2022cqj}
J.~P.~Lees \textit{et al.} [BaBar],
\emph{Search for Heavy Neutral Leptons Using Tau Lepton Decays at BABAR},
[arXiv:2207.09575 [hep-ex]].

\bibitem{Kim:2018uht}
C.~S.~Kim, G.~L\'opez Castro and D.~Sahoo,
\emph{Constraints on a sub-eV scale sterile neutrino from nonoscillation measurements} (2018),
Phys. Rev. D \textbf{98} (2018) no.11, 115021.

\bibitem{47}
 M. Fael, L. Mercolli and M. Passera \emph{W-propagator corrections to $\mu$ and $\tau$ leptonic decays} (2013),     Phys.Rev.D 88 (2013) 9, 093011.
 
 \bibitem{Bodrov:2022mbd}
D.~Bodrov and P.~Pakhlov,
\emph{A new method for measurement of the Michel parameters that describe the daughter muon polarization in the $\tau^- \to \mu^- \bar{\nu}_\mu\nu_\tau$ decay},
[arXiv:2203.12743 [hep-ph]].
 
 \bibitem{48}
 Alberto Sirlin and Andrea Ferroglia, \emph{Radiative Corrections in Precision Electroweak Physics: Historical Perspective} (2012),     Rev.Mod.Phys. 85 (2013) 1, 263-297.
 
 \bibitem{49}
 Boris Kayser, \emph{Majorana neutrinos and their electromagnetic properties} (1982),     Phys.Rev.D 26 (1982) 1662. 
 
 \bibitem{50}
C.S. Kim, M.V.N. Murthy, and Dibyakrupa Sahoo, \emph{Inferring the nature of active neutrinos: Dirac or Majorana?},  
Phys. Rev. D {\bf105} (2022) 113006.




\bibitem{53}
Andrea Ferroglia, Christoph Greub, Alberto Sirlin and Zhibai Zhang, \emph{Contributions of the W-boson propagator to the $\mu$ and $\tau$ leptonic decay rates},     Phys.Rev.D 88 (2013) 3, 033012.


\bibitem{vanRitbergen:1998yd}
T.~van Ritbergen and R.~G.~Stuart,
\emph{Complete two loop quantum electrodynamic contributions to the muon lifetime in the Fermi model},
Phys. Rev. Lett. \textbf{82} (1999), 488-491.

\bibitem{vanRitbergen:1999fi}
T.~van Ritbergen and R.~G.~Stuart,
\emph{On the precise determination of the Fermi coupling constant from the muon lifetime},
Nucl. Phys. B \textbf{564} (2000), 343-390.

\bibitem{Caola:2014daa}
F.~Caola, A.~Czarnecki, Y.~Liang, K.~Melnikov and R.~Szafron,
\emph{Muon decay spin asymmetry},
Phys. Rev. D \textbf{90} (2014) no.5, 053004.

\bibitem{Fael:2020tow}
M.~Fael, K.~Sch\"onwald and M.~Steinhauser,
\emph{Third order corrections to the semileptonic b\textrightarrow{}c and the muon decays},
Phys. Rev. D \textbf{104} (2021) no.1, 016003

\bibitem{Czakon:2021ybq}
M.~Czakon, A.~Czarnecki and M.~Dowling,
\emph{Three-loop corrections to the muon and heavy quark decay rates},
Phys. Rev. D \textbf{103} (2021), L111301.

\bibitem{Czarnecki:2014cxa}
A.~Czarnecki, M.~Dowling, X.~Garc\'ia Tormo, i, W.~J.~Marciano and R.~Szafron,
\emph{Michel decay spectrum for a muon bound to a nucleus},
Phys. Rev. D \textbf{90} (2014) no.9, 093002.

\bibitem{55}
P. Langacker, \emph{The Standard Model and Beyond}, CRC Press (2010).
 
\bibitem{56}
Palash B. Pal, \emph{An Introductory course of particle physics}, CRC Press (2015).

\bibitem{57}
K. Mursula and F.Scheck, \emph{Analysis of leptonic charged weak interactions}, Nuclear Physics {\bf B253} (1985).

\bibitem{58}
W. Fetscher, H.J. Gerber and K.F. Johnson, \emph{Electroweak and Strong Interactions}, Springer Verlag (1996).

\bibitem{Belle-II:2018jsg}
E.~Kou \textit{et al.} [Belle-II],
\emph{The Belle II Physics Book},
PTEP \textbf{2019} (2019) no.12, 123C01
[erratum: PTEP \textbf{2020} (2020) no.2, 029201]

\bibitem{Eidelman:2015wja}
S.~Eidelman, \emph{Project of the Super-tau-charm Factory in Novosibirsk},
Nucl. Part. Phys. Proc. \textbf{260} (2015), 238-241
doi:10.1016/j.nuclphysbps.2015.02.050.

\bibitem{Epifanov:2017kly}
D.~A.~Epifanov [Belle],
\emph{Measurement of Michel parameters in leptonic \ensuremath{\tau} decays at Belle},
Nucl. Part. Phys. Proc. \textbf{287-288} (2017), 7-10.

\bibitem{Bodrov:2021hfe}
D.~A.~Bodrov,
\emph{Measurement of the Michel Parameter ${\xi^{\prime}}$ in Tau-Lepton Decays at the Super Charm-Tau Factory},
Phys. Atom. Nucl. \textbf{84} (2021) no.2, 212-215.

\bibitem{Lyu:2021tlb}
X.~R.~Lyu [STCF Working Group],
\emph{Physics Program of the Super Tau-Charm Factory},
PoS \textbf{BEAUTY2020} (2021), 060
.

\bibitem{Aiba:2021bxe}
M.~Aiba, A.~Amato, A.~Antognini, S.~Ban, N.~Berger, L.~Caminada, R.~Chislett, P.~Crivelli, A.~Crivellin and G.~D.~Maso, \textit{et al.}
\emph{Science Case for the new High-Intensity Muon Beams HIMB at PSI},
[arXiv:2111.05788 [hep-ex]].






\end{thebibliography}
\end{document}